\documentclass[fleqn,usenatbib,useAMS]{mnras}

\usepackage[dvipsnames]{xcolor}
\usepackage{tikz,hyperref}

\definecolor{lime}{HTML}{A6CE39}
\DeclareRobustCommand{\orcidicon}{
	\begin{tikzpicture}
	\draw[lime, fill=lime] (0,0) 
	circle [radius=0.14] 
	node[white] {{\fontfamily{qag}\selectfont \tiny ID}};
	\draw[white, fill=white] (-0.0625,0.000) 
	circle [radius=0.007];
	\end{tikzpicture}
	\hspace{-2mm}
}

\foreach \x in {A, ..., Z}{\expandafter\xdef\csname orcid\x\endcsname{\noexpand\href{https://orcid.org/\csname orcidauthor\x\endcsname}
			{\noexpand\orcidicon}}
}

\usepackage{graphicx}	
\usepackage{amsmath}	
\usepackage{amssymb,pifont}	
\newcommand{\xmark}{\ding{55}}
\newcommand{\cmark}{\ding{51}}
\usepackage{multicol}        
\usepackage{bm}		
\usepackage{pdflscape}	
\usepackage[encapsulated]{CJK}
\usepackage{ucs}
\usepackage[utf8x]{inputenc}
\usepackage{multirow}
\usepackage{booktabs,caption}
\usepackage[flushleft]{threeparttable}
\usepackage[dvipsnames]{xcolor}
\usepackage{adjustbox}

\usepackage{rotating} 
\usepackage{lscape}

\newcommand{\hi}{H\,{\scshape i}}

\newcommand{\ha}{H$\alpha$}
\newcommand{\hb}{H$\beta$}

\newcommand{\hei}{He{\,\scshape i}}
\newcommand{\heii}{He{\,\scshape ii}}
\newcommand{\cii}{C{\,\scshape ii}}
\newcommand{\ciii}{C{\,\scshape iii}}
\newcommand{\civ}{C{\,\scshape iv}}
\newcommand{\cliv}{[Cl{\,\scshape iv}]}

\newcommand{\nii}{[N{\,\scshape ii}]}
\newcommand{\niii}{N{\,\scshape iii}}

\newcommand{\nv}{N{\,\scshape v}}
\newcommand{\ovi}{O{\,\scshape vi}}

\newcommand{\ov}{O{\,\scshape v}}
\newcommand{\oiii}{[O{\,\scshape iii}]}
\newcommand{\oii}{[O{\,\scshape ii}]}
\newcommand{\oi}{[O{\,\scshape i}]}

\newcommand{\sii}{[S{\,\scshape ii}]}
\newcommand{\siii}{[S{\,\scshape iii}]}

\newcommand{\cliii}{[Cl{\,\scshape iii}]}

\newcommand{\ariii}{[Ar{\,\scshape iii}]}
\newcommand{\ariv}{[Ar{\,\scshape iv}]}

\newcommand{\neiii}{[Ne{\,\scshape iii}]}

\usepackage[T1]{fontenc}
\usepackage{ae,aecompl}

\usepackage{newtxtext,newtxmath}

\title[PNe with WR-CSPN - IV]{Planetary nebulae with Wolf-Rayet-type central stars - IV. NGC\,1501 and its mixing layer}
\author[G.\ Rubio et al.]{G.\,Rubio\thanks{E-mail: grubio@idec.edu.mx}$^{1,2\orcidA}$, J.~A.\,Toal\'{a}$^{3\orcidB}$, H.\,Todt$^{4}$, L.\,Sabin$^{5\orcidE}$, E.\,Santamar\'{i}a$^{1,2\orcidD}$, G.\,Ramos-Larios$^{1,2\orcidC}$ and 
\newauthor{M.~A.\,Guerrero$^{6}\orcidF$}\\
$^1$CUCEI, Universidad de Guadalajara, Blvd. Marcelino Garc\'\i a Barrag\'an 1421, 44430, Guadalajara, Jalisco, Mexico \\
$^2$Instituto de Astronom\'\i a y Meteorolog\'\i a, Dpto.\ de F\'\i sica, CUCEI, Av.\ Vallarta 2602, 44130, Guadalajara, Jalisco, Mexico\\
$^3$Instituto de Radioastronom\'{i}a y Astrof\'{i}sica, UNAM, Antigua Carretera a P\'{a}tzcuaro 8701, Ex-Hda. San Jos\'{e} de la Huerta, Morelia 58089, Mich., Mexico\\
$^4$Institute for Physics and Astronomy, Universit\"{a}t Potsdam, Karl-Liebknecht-Str 24/25, D-14476 Potsdam, Germany\\
$^{5}$Instituto de Astronom\'{i}a, UNAM, Apdo. Postal 877, Ensenada 22860, B.C., Mexico\\
$^{6}$Instituto de Astrof\'{i}sica de Andaluc\'{i}a, IAA-CSIC, Glorieta de la Astronom\'{i}a S/N, Granada 18008, Spain}

\pubyear{2022}

\begin{document}
\label{firstpage}
\pagerange{\pageref{firstpage}--\pageref{lastpage}}
\maketitle

\begin{abstract}
Theory predicts that the temperature of the X-ray-emitting gas ($\sim$10$^{6}$~K)  detected from planetary nebulae (PNe) is a consequence of mixing or thermal conduction when in contact with the ionized outer rim ($\sim$10$^{4}$~K). Gas at intermediate temperatures ($\sim$10$^{5}$~K) can be used to study the physics of the production of X-ray-emitting gas, via C\,{\sc iv}, N\,{\sc v} and O\,{\sc vi} ions. Here we model the stellar atmosphere of the CSPN of NGC\,1501 to demonstrate that even this hot H-deficient [WO4]-type star cannot produce these emission lines by photoionization. We use the detection of the C\,{\sc iv} lines to assess the physical properties of the mixing region in this PNe in comparison with its X-ray-emitting gas, rendering NGC\,1501 only the second PNe with such characterization. We extend our predictions to the hottest [WO1] and cooler [WC5] spectral types and demonstrate that most energetic photons are absorbed in the dense winds of [WR] CSPN and highly ionized species can be used to study the physics behind the production of hot bubbles in PNe. We found that the UV observations of NGC\,2452, NGC\,6751 and NGC\,6905 are consistent with the presence mixing layers and hot bubbles, providing excellent candidates for future X-ray observations.
\end{abstract}

\begin{keywords}
stars: evolution --- stars: winds, outflows --- stars: Wolf-Rayet --- stars: individual: WD\,0402+607 --- (ISM:) planetary nebulae: general --- (ISM:) planetary nebulae: individual: NGC\,1501
\end{keywords}




\section{INTRODUCTION}
\label{sec:intro}

Planetary nebulae (PNe) form as the result of mass loss experienced during the evolution of low- and intermediate-mass stars (1$\lesssim M_\mathrm{i}$/M$_{\odot} \lesssim$8). These stars evolve through the asymptotic giant branch (AGB) phase exhibiting slow and dense
winds \citep[$v_\mathrm{AGB}\approx20$~km~s$^{-1}$, $\dot{M}_\mathrm{AGB}\lesssim10^{-5}$~M$_\odot$~yr$^{-1}$;][]{Ramstedt2020}. After
exposing their hot cores, these stars evolve into the post-AGB phase increasing their temperatures and developing a tenuous but fast stellar wind \citep[$v_{\infty} \gtrsim$ 1000~km~s$^{-1}$;][]{Guerrero2013}. The former pushes and compresses the AGB material into a dense shell that expands into the surrounding medium. Simultaneously, the progenitor star develops a strong UV flux which ionizes this shell. The combination of these processes create what we know as PNe \citep{Kwok2000}.

Currently, we know that this simplistic scenario is insufficient to explain the variety of complex structures and morphologies of all PNe. 
It has been accepted that binarity plays a major role shaping PNe \citep[e.g.,][]{DeMarco2009}, with the common envelope channel being the most promising one \citep{Ivanova2013}. Recent searches for binary central stars of PNe (CSPNe) have been pushing the limits of our understanding \citep[see, e.g.,][]{Jacoby2021,chornay2021}, in addition to the new generation of instruments \citep[e.g.,][]{GarciaRojas2022,Rechy2022} and computational tools \citep[see][and references therein]{Ondratschek2022}.

Regardless of the details, the fast-wind-slow-wind interaction produces an adiabatically-shocked region filling the ionized structure of a PN. Theoretically, this shock produces a hot bubble with a temperature that depends on the stellar wind velocity as \citep[see, e.g.,][]{Dyson1997}
\begin{equation}
    T = \frac{3}{16} \frac{\overline{m}}{k_\mathrm{B}} v_{\infty}^{2},
\end{equation}
\noindent where $k_\mathrm{B}$ is the Boltzmann constant and $\overline{m}$ is the mean particle mass. For a typical fast wind should be expected to reach values in the range of 10$^{7}$--10$^{8}$~K. In stark contrast with those theoretical predictions X-ray observations of PN report hot bubble temperatures of a few times 10$^{6}$~K \citep[see][and references therein]{Chu2001,Toala2020,Ruiz2013,Yu2009}.

It has been long argued that the temperature discrepancy is due to extra physics acting at the outer edge of the hot bubble, the region in contact with the nebular material ($\sim$10$^{4}$~K). The most common physical process is thermal conductivity \citep[e.g.,][]{Soker1994,Sandin2016}, where electrons from the hot bubble evaporate material from the photoionized nebula. However, multi-dimension numerical simulations have shown that the wind-wind interaction region corrugates under Rayleigh-Taylor and thin-shell instabilities producing a set of clumps and filaments that naturally mix the ionized region with the hot bubble \citep[e.g.,][]{Stute2006,Toala2016}. In both scenarios, the inclusion of material reduces the temperature of the hot bubble at its outer edge while raising the density. As a consequence, the X-ray emissivity of the soft X-ray band increases to detectable limits \citep[e.g.,][]{Steffen2008,Toala2018}.

Most of the hot bubbles detected through X-ray observations correspond to PNe harbouring H-deficient  CSPN, in particular those classified as [Wolf-Rayet] ([WR]) type \citep[][]{Kastner2012,Freeman2014,Toala2019a}. Although the reason is not completely clear, one might suggest that is the result of the powerful winds from [WR]-type CSPNe interacting with the slow AGB material producing higher degree of turbulent structures \citep{Gesicki2006,Medina2006}.

\begin{figure}
\begin{center}
  \includegraphics[width=0.97\linewidth]{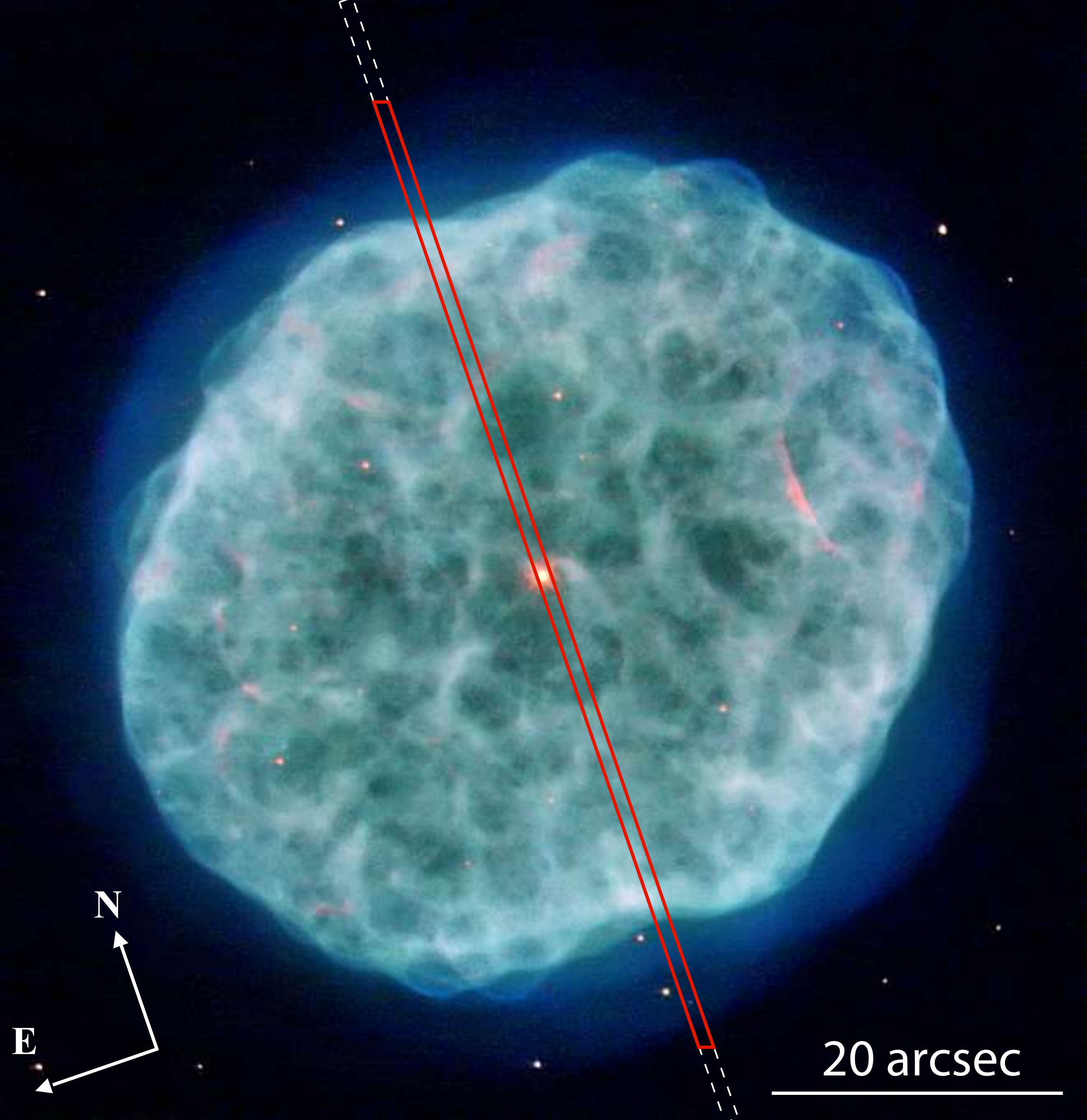}
\caption{Colour composite {\it HST} image of NGC\,1501. Red, green and blue correspond to the [N\,{\sc ii}], H$\alpha$ and [O\,{\sc iii}] narrow band filter images, respectively, obtained from the Hubble Legacy Archive (\url{https://hla.stsci.edu/}). The WHT ISIS slit position is shown with a (white) dashed line polygonal aperture and has a PA=0$^\circ$.The red rectangle represents the spectrum extraction region.}
\label{fig:HST}
\end{center}
\end{figure}

Independently of the physical process acting at the edge of the hot bubble, a conductive or mixing layer is expected to appear at the interface between the hot bubble ($\sim$10$^{6}$~K) and the optically-emitting nebular material ($\sim$10$^{4}$~K) with intermediate temperatures in the 10$^{5}$~K range. Following this criteria, the presence of highly ionized species such as C\,{\sc iv}, N\,{\sc v} and O\,{\sc vi} at UV wavelengths have been suggested as probes of the presence of X-ray emission from hot bubbles \citep{Iping2002,Gruendl2004}. These ions have ionization potentials of 47.9, 77.5, and 113.9~eV and have fractional abundances peaking between [1--3]$\times10^{5}$~K \citep{Shull1982}. \citet{Ruiz2013} argued that C\,{\sc iv}, N\,{\sc v} and O\,{\sc vi} can be easily produced via photoionization by CSPN with effective temperatures ($T_\mathrm{eff}$) larger than $\sim$35,000, $\sim$80,000 and $\sim$140,000~K, respectively, assuming that the CSPN emit blackbody radiation. Then, \citet{Ruiz2013} confirmed that the detection of O\,{\sc vi} surrounding CSPN with low $T_\mathrm{eff}$ can be related to the presence of hot bubbles in PNe. More recently, \citet{Fang2016} used {\it Hubble Space Telescope} ({\it HST}) Space Telescope Imaging Spectrograph observations to show that the N\,{\sc v} line traces the mixing layer in the X-ray-emitting PN NGC\,6543, which harbours a CSPN with $T_\mathrm{eff}$=50,000~K.
However, it must be noted that the situation with CSPN of the [WR]-type is different. Most of the radiation produced by [WR] CSPNe is absorbed by their massive winds. As a consequence, the effect of their ionization photon flux is effectively smaller than the case of H-rich CSPNe.

To assess these effects, we use publicly available observations of the [WR] PN (hereinafter WRPN) NGC\,1501 around the [WR]-type WD\,0402+607 (a.k.a. V* CH Cam)\footnote{We will use the white dwarf designation (WD\,0402+607) for this CSPN throughout the paper, but we warn the reader that its [WR] nature has been widely known \citep[see, e.g.,][]{Acker2003}.}. 
NGC\,1501 is a high-excitation WRPN with a relatively round clumpy structure (see Fig.~\ref{fig:HST}). Its ionization structure and kinematics have been studied in the past \citep[see][and references therein]{Sabbadin2000,Ragazzoni2001}. The most complete work addressing the physical properties and abundances of NGC\,1501 was presented by \citet{Ercolano2004}. These authors used high-resolution spectroscopic observations to study its abundances and ionization structure as well as to characterize the properties of its [WR]-type CSPN. \citet{Ercolano2004} constructed a 3D photoionization model of NGC\,1501 to explore the physical properties of the collisionally excited and optical recombination lines. Of particular interest, these authors reported the presence of the C\,{\sc iv}~5801.51,5812.14~\AA\, doublet and the N\,{\sc v}~7618.5~\AA\, emission lines in the nebular spectrum of NGC\,1501, but did not further discuss their origin. Given that NGC\,1501 has been found to emit extended X-rays using the {\it Chandra X-ray Observatory} \citep{Freeman2014}, the presence of such emission lines would suggest the presence of a mixing layer within this WRPN.

In this paper we present an updated model of NGC\,1501 and its CSPN to assess the formation of emission lines in the mixing region. We model the broad spectral features of WD\,0402+607 to produce a synthetic model of its stellar atmosphere. This stellar atmosphere model is subsequently used to create a photoionization model including the contribution from gas and dust in NGC\,1501. We extend our calculations to cover the range of the hottest [WO]-type stars to the cooler [WC5]-type and make some predictions on the possible X-ray emission from other WRPNe.

\begin{figure*}
\begin{center}
  \includegraphics[width=0.95\linewidth]{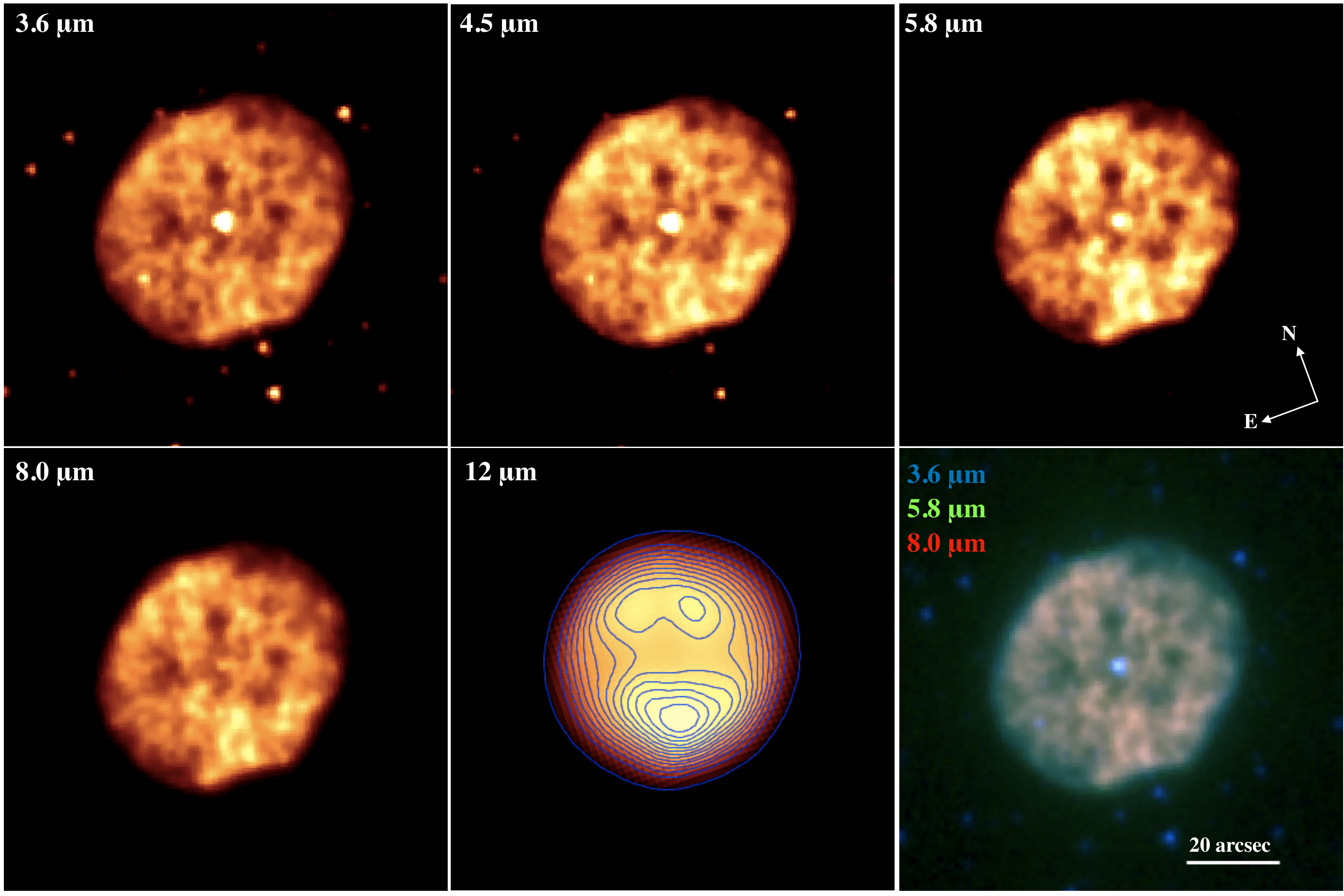}
\caption{IR views of NGC\,1501. 
The images correspond to the {\it Spitzer} IRAC images at 3.6, 4.5, 5.8, 8.0 $\mu$m  while the 12~$\mu$m image corresponds to the {\it WISE} band 3, all presented in their native spatial resolution. The contours on the {\it WISE} image are only used to emphasize the spatial distribution of the emission in this band. The bottom right panel shows a colour-composite image of NGC\,1501 obtained by combining the {\it Spitzer} 3.6 (blue), 5.8 (green) and 8.0 (red) bands. 
All panels have the same field of view.}
\label{fig:IR_views}
\end{center}
\end{figure*}

\section{Observations and data preparation}
\label{sec:observations}

In this work we produced a fit to the observables of the nebular and dust properties of NGC\,1501 as well as the atmosphere of its CSPN. For this, we retrieved a set of publicly available observations covering the UV, optical, IR and radio wavelengths. Here we give some details of the archival observations used in this paper and their preparation.

\subsection{Optical spectroscopic observations}

To fully characterize NGC\,1501 and its [WR]-type CSPN, \citet{Ercolano2004} analysed Intermediate-dispersion Spectrograph and Imaging System (ISIS) observations obtained at the 4.2~m William Herschel Telescope (WHT) at the Observatorio Roque de los Muchachos (La Palma, Spain). The WHT ISIS observations were carried out on 2003 August 1 (PI: R.\,Wesson; Prop.\,ID. W2003B/62). The WHT ISIS observations were retrieved from the Isaac Newton Group Archive\footnote{\url{http://casu.ast.cam.ac.uk/casuadc/ingarch/query}}. These are still the deepest optical spectroscopic observations of NGC\,1501 and will be used in the present paper.

The WHT ISIS observations were obtained with the R316R and R600B gratings which cover the red (5100--8000~\AA) and blue (3400-- 5100~\AA) wavelength ranges, respectively. In both cases the slit was positioned on WD\,0402$+$607 oriented N-S, that is, with a position angle (PA) of 0$^{\circ}$ as illustrated in Figure~\ref{fig:HST}. A slit width of 0.75~arcsec was used which provided spectral resolutions of 1.5 and 2.8~\AA\, for the blue and red arms, respectively. The total observing time on each arm was 9000~s. The WHT ISIS data were reduced following {\sc iraf} standard routines \citep{Tody1993}. The resultant spectra of NGC\,1501 and WD\,0402+607 (not shown here) are almost identical to those shown in \citet{Ercolano2004}.

\subsection{IR observations}

To construct a consistent model of the nebular properties of NGC\,1501 we also need to include the contribution from dust as photons not only interact with the gas but also with the dust present in the nebula \citep[e.g.,][]{GLL2018,Toala2021}. IR observations of NGC\,1501 from different missions were obtained from the NASA/IPAC Infrared Science Archive (IRSA)\footnote{\url{https://irsa.ipac.caltech.edu/frontpage/}}. We collected IR data of NGC\,1501 that cover the 3--160~$\mu$m wavelength range.

NGC\,1501 was observed by the {\it Spitzer Space Telescope} (hereinafter {\it Spitzer}) with the Infrared Array Camera \citep[IRAC;][]{Fazio2004} on 2007 October 16 with the 3.6, 4.5, 5.8 and 8.0~$\mu$m bands. The {\it Spitzer} IRAC observations were obtained as part of the observation ID 40115 (PI: G.\,Fazio) and correspond to the AORKey 21971712. The {\it Spitzer} IRAC images are shown in Fig.~\ref{fig:IR_views}.

Observations from the {\it Wide-field Infrared Survey Explorer (WISE)} at 3.4, 4.5, 12 and 22~$\mu$m and {\it Infrared Astronomical Satellite (IRAS)} at 25, 60 and 100~$\mu$m were also retrieved from the IRSA. The {\it WISE} band 3 (12~$\mu$m) image is compared to the {\it Spitzer} IRAC images in Fig.~\ref{fig:IR_views}, but longer wavelength IR images do not properly resolve the structure of NGC\,1501 and are not shown here. The IRAC IR images presented in Fig.~\ref{fig:IR_views} resemble the clumpy structure unveiled by the higher resolution {\it HST} image presented in Fig.~\ref{fig:HST}.

\begin{figure}
\begin{center}
\includegraphics[width=0.98\linewidth]{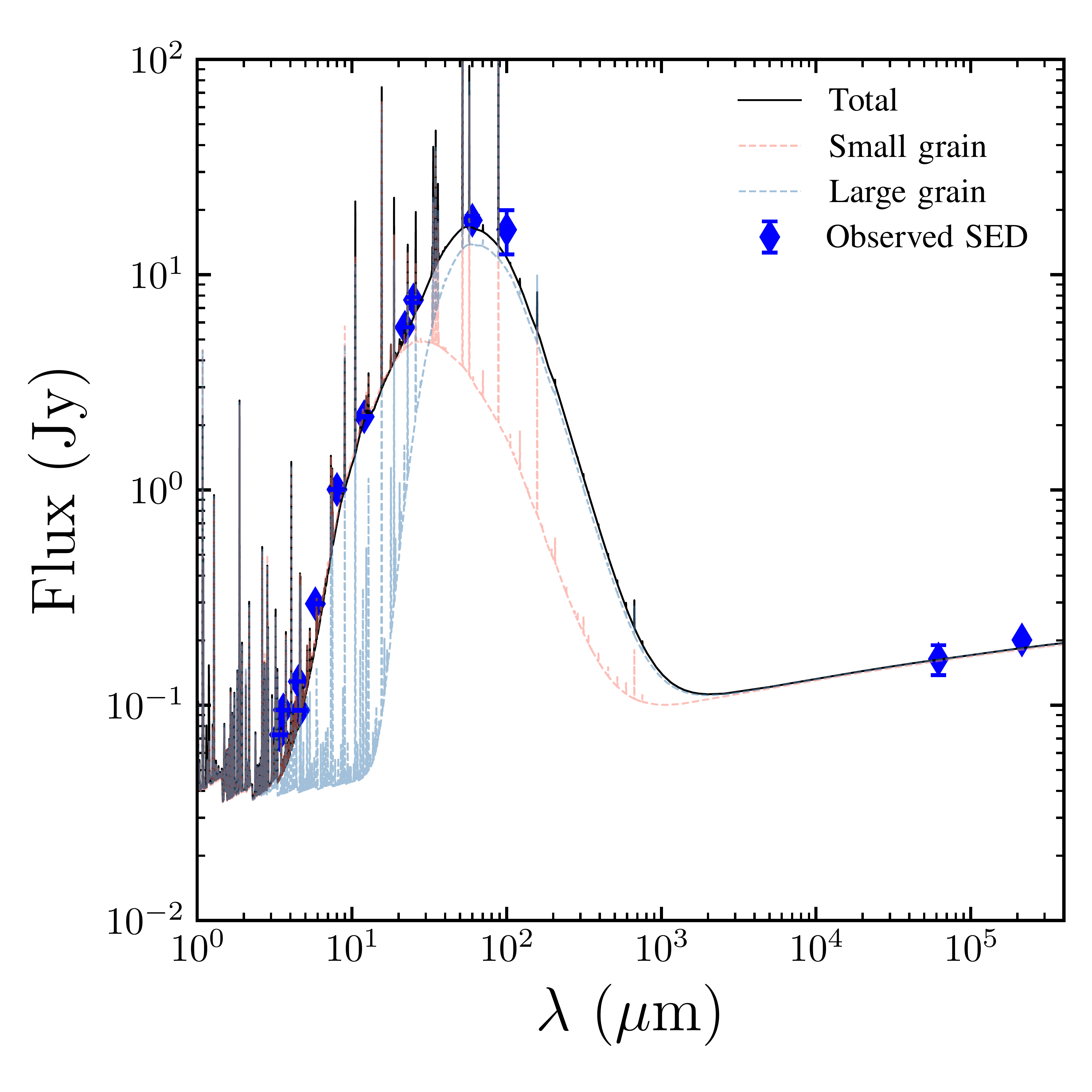}
\caption{Observed IR and radio SED of NGC\,1501 (blue diamonds with error bars). The black solid line shows the total synthetic spectrum obtained from our best {\scshape cloudy} model.}
\label{fig:sed_cont}
\end{center}
\end{figure}

To construct a spectral energy distribution (SED), we extracted photometric fluxes and errors from the IR observations following the procedure described in the recent papers published by our group \citep[see][]{Rubio2020,JH2020,JH2021,GG2022}. In addition, we also adopted the 1.4 and 4.85~GHz radio measurements reported in \citet{Hajduk2018} obtained from Very Large Array (VLA) observations of the National Radio Astronomy Observatory (NRAO) and the NRAO Green Bank Telescope (GBT). All photometric values and errors are listed in Table~\ref{tab:ir_photometry} and are plotted in Fig.~\ref{fig:sed_cont}.

\begin{table}
\begin{center}
\caption{IR and radio photometric fluxes and errors of NGC\,1501.}
\setlength{\tabcolsep}{\tabcolsep}    
\begin{tabular}{lccc}
\hline
Wavelength & Band     & Flux  & Error\\
range      & ($\mu$m) & (mJy) & (mJy)\\
\hline     
IR         & 3.4      & 73    & 1 \\
           & 3.6      & 95   & 1 \\
           & 4.5       & 129   & 1 \\
           & 4.6       & 95  & 1 \\
           & 5.8       & 296  & 3 \\
           & 8.0      & 1005  & 7 \\
           & 12      & 2189  & 17 \\
           & 22      & 5696  & 6 \\
           & 25      & 7630 & 220  \\
           & 60      & 17900 & 840 \\     
           & 100      & 16180  & 3760 \\
\hline
Radio      & 61813    & 164    & 26 \\
           & 214137   & 201    & 7 \\
\hline
\end{tabular} 
\label{tab:ir_photometry}
\end{center}
\end{table}

\begin{figure*}
\begin{center}
\includegraphics[width=0.85\linewidth]{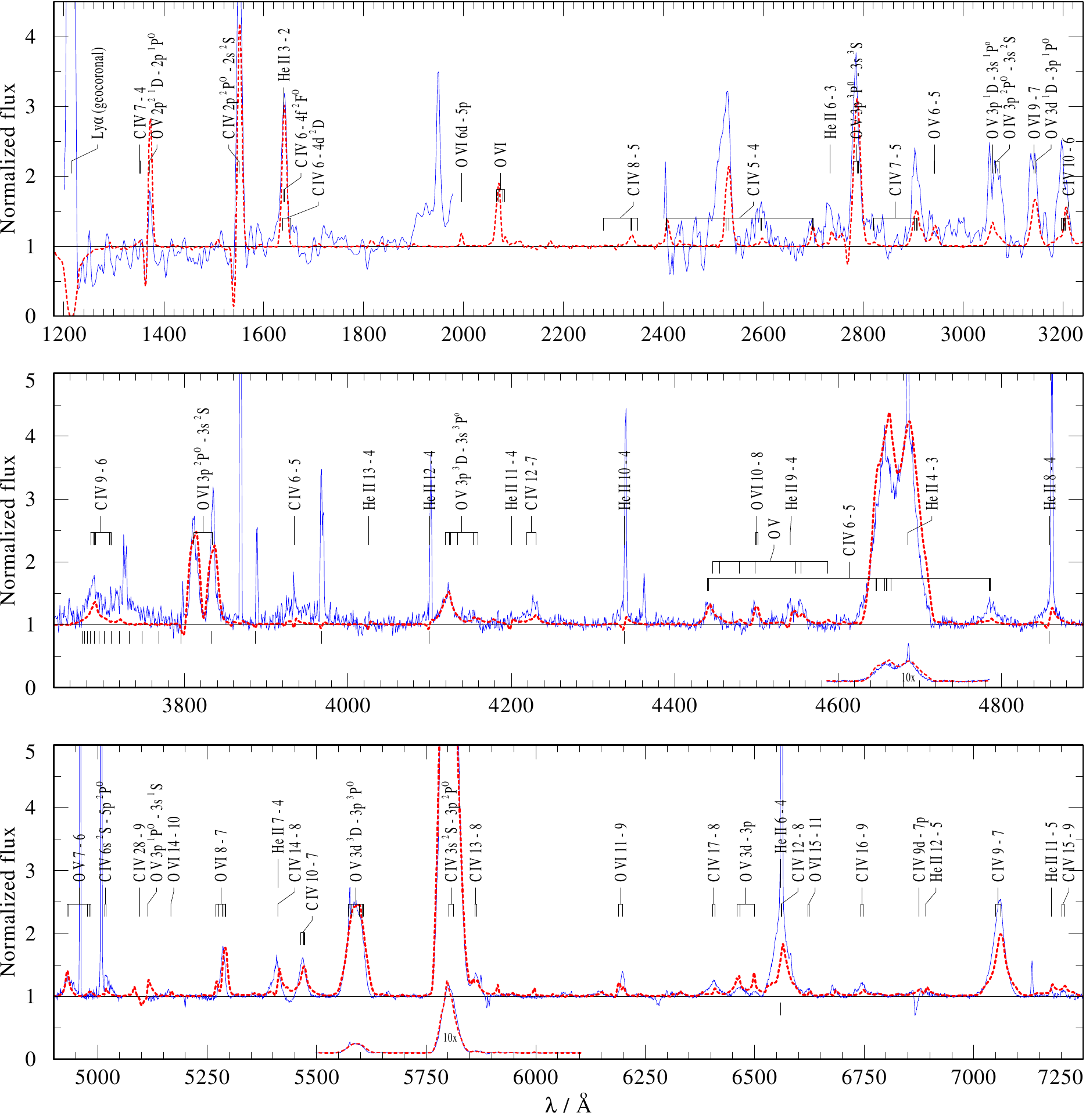}
\caption{Comparison between our {\sc powr} NLTE stellar atmosphere model (red dashed line) and observations of the CSPN (blue line). 
The upper panel shows a comparison with UV {\itshape IUE} data, where the model spectrum has been convolved with a Gaussian of 5\,\AA{} (FWHM) to match the resolution of the low dispersion observation. The middle and bottom panels show a comparison with the \textit{WHT ISIS} data. The short-wavelength range of the \textit{ISIS} observations is binned to 0.5\,\AA{} for a better representation. The
 black lines below the continuum in the 2nd and 3rd panel mark the nebular Balmer lines.}
\label{fig:PoWR}
\includegraphics[width=0.85\linewidth]{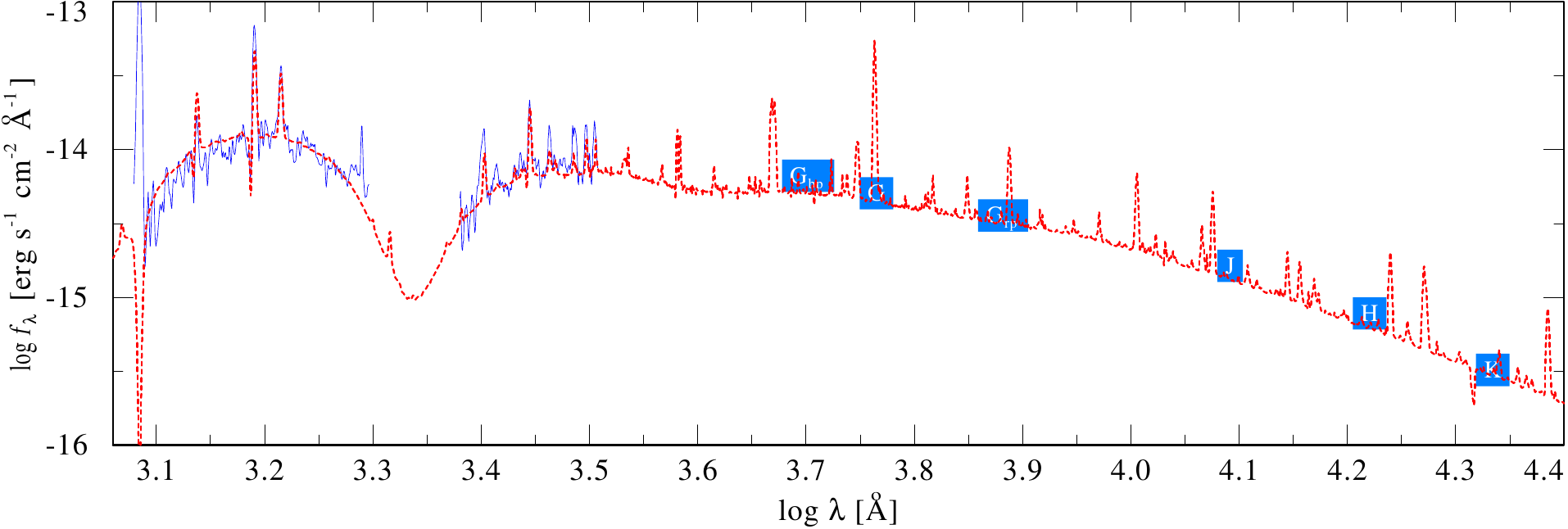}
\caption{SED of the CSPN of NGC\,1501: Observations (blue solid lines and boxes) vs.\ best fitting {\sc powr} model (red dashed line). The plots shows the \textit{IUE} (short and long wavelengths) spectroscopic observations and \textit{Gaia} and \textit{2MASS} photometric data.}
\label{fig:SED_POWR}
\end{center}
\end{figure*}

\begin{figure}
\begin{center}
  \includegraphics[width=\linewidth]{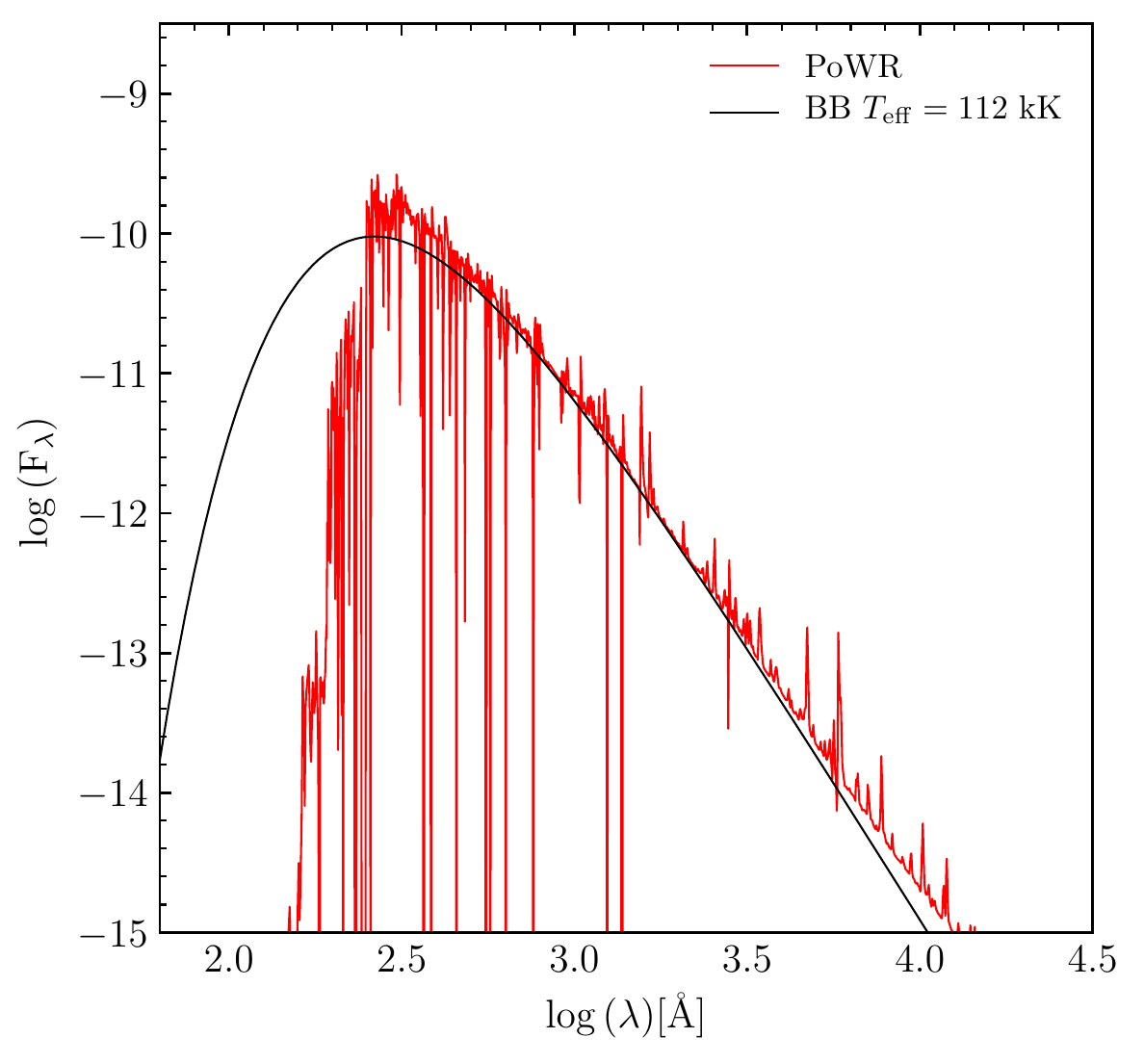}
\caption{Comparison between the best fit {\sc powr} stellar atmosphere model of the CSPN of NGC\,1501 and a black body with a similar $T_{\mathrm{eff}}$ of 112~kK.}
\label{fig:SED}
\end{center}
\end{figure}

\subsection{UV data}

To produce a model of the stellar atmosphere of the CSPN of NGC\,1501, UV data from the {\itshape Far Ultraviolet Spectroscopic Explorer} ({\itshape FUSE}) and {\itshape International Ultraviolet Explorer} ({\itshape IUE}) were retrieved from the Mikulski Archive for Space Telescopes (MAST)\footnote{\url{https://archive.stsci.edu/hst/}}. The {\itshape FUSE} data correspond to the Obs.\,ID Z9110601000 with a total observing time of 15502.9~s. Unfortunately, the \textit{FUSE} observation seems to be totally underexposed and can therefore not be used for the analysis.  
The {\itshape IUE} observations correspond to the Obs.\,IDs LWP08948, SWP28952, SWP28953 and SWP44229 with exposure times of 9000, 3300, 9960 and 13200~s, respectively.

\section{NLTE analysis of the central star}
\label{sec:PoWR}

\citet{Ercolano2004} presented a thorough model of the ionization structure of NGC\,1501. They used a stellar atmosphere model computed with the T\"{u}bingen NLTE Model Atmosphere Package \citep{Werner2012} for plane-parallel atmospheres, where $T_\mathrm{eff}$ and gravity ($g$) were varied to provide an ionizing spectrum that could reproduce the observed ionization structure.
The stellar parameters resulted in $T_\mathrm{eff}$=110~kK, log($g$)=6.0 (in cgs units), and $L_\star$=5000~L$_{\odot}$, assuming a distance of 1.30\,kpc. Moreover
they used a recombination line analysis to derive the stellar wind abundances of He:C:N:O=33:50:2:15 by mass.

In this paper we model the optical and available UV spectra of the central star using the most recent version\footnote{2022 February 21} of the non-local thermodynamic equilibrium (NLTE) Potsdam stellar atmosphere code \citep[{\sc powr};][]{Grafener2002,Hamann2004}\footnote{\url{https://www.astro.physik.uni-potsdam.de/PoWR}}. Details on the code performance can be found in \citet{Todt2015} and applications to WR-type CSPNe can be found in our group's recent papers \citep[see][]{GG2020,GG2022,Toala2019}.

\begin{table}
\caption{Stellar parameters of our best {\sc powr} model for the CSPN of NGC\,1501.}
\label{tab:powrparameters}
\begin{center}
\begin{tabular}{lll}
\hline
parameter & value & comment \\
\hline
$d$ [kpc]    & $1.64\pm0.06$       & \cite{BJ2021} \\
$E_{B-V}$ [mag] & $0.8_{-0.02}^{+0.04}$ \\
$M_{\star}$ [M$_\odot$]              & 0.6               & adopted \\
$T_\star$ [kK] & $112\pm2$ &  \\
$\log(L_\star/L_\odot)$ & $4.1\pm0.1$ & \\
$R_\star$ [R$_\odot$]  &   $0.29\pm0.01$ & \\
$R_\mathrm{t}$ [R$_\odot$] & $9.1\pm0.7$ & transformed radius\\
$\log \dot{M}$ [M$_\odot$~y$^{-1}$] &  $-6.8\pm0.2$ & \\
$\varv_\infty$ [km/s] & $2000\pm200$ &\\
$D$                                  & 10                & density contrast \\
$\beta$                              & 1                 & $\beta$-law exponent\\
\hline
\multicolumn{3}{l}{Chemical abundances (mass fraction)}\\
\hline
$X_\text{He}$ & $0.60\pm0.10$ & \\
$X_\text{C}$ & $0.30\pm0.10$ &\\
$X_\text{O}$ & $0.15\pm0.05$ &  \\
$X_\text{Fe}$ & 0.0014 & adopted, iron group elements \\
\hline
\end{tabular}
\end{center}
\end{table}

The emission-line spectra of WR-type stars are mainly formed by recombination processes in their dense
stellar winds. Hence, the continuum-normalized spectrum
shows a useful scale-invariance: for a given stellar temperature ($T_\star$) 
and chemical composition, the equivalent width of the emission lines depend 
in first approximation only on the volume emission measure of the wind normalized to the area of
the stellar surface. A corresponding quantity is the transformed radius $R_\mathrm{t}$, 
which has been introduced by \citet{Schmutz1989} and is defined as:
\begin{equation}
    R_\mathrm{t} = R_{\star} \left[\frac{\varv_{\infty}}{2500~\mathrm{km}~\mathrm{s}^{-1}} \bigg/ \frac{\dot{M} \sqrt{D}}{10^{-4}\, \mathrm{M}_{\odot}~\mathrm{yr}^{-1}}\right]^{2/3}~.  
\label{eq:r_t}
\end{equation}
Therefore, different combinations of mass-loss rates ($\dot{M}$) and stellar radii ($R_\star$) 
can lead to the same strength of the emission-lines. This invariance also includes the micro-clumping
parameter $D$, which is defined as the density contrast between wind clumps and a smooth wind of the same $\dot{M}$. Hence, empirically derived $\dot{M}$ depend on the adopted value of $D$.
The parameters of our best fitting model for the stellar atmosphere of the CSPN obtained with {\sc powr} are listed in Table~\ref{tab:powrparameters} and the best fitting model is shown in Fig.~\ref{fig:PoWR} in comparison with the observed UV and optical data. We note that the fit quality to the UV spectra is acceptable, taking into account the limited quality of the \textit{IUE} spectra. 

For our calculations we adopt a stellar mass of $M_{\star}$=0.6~M$_{\odot}$, which is the typical value for an evolved low-mass star \citep[see][and references therein]{MillerBertolami2016,Kepler2016}. We note that the actual mass has little impact on the spectrum of the CSPN, given that the spectrum is formed in the wind.

We obtained an initial guess for the parameters $T_\star$ and $R_\text{t}$ from calculating a reduced $\chi^2$ for a grid of [WCE] models (with fixed chemical composition, terminal velocity, stellar luminosity, etc.). These parameters are then further refined by calculating individual models. For these models also the chemical composition, stellar luminosity, etc., are varied to achieve a sufficient fit. The fit quality is evaluated by eye, where a weight is put on sensitive lines. The uncertainties of the inferred parameters give the parameter range for which the model spectrum is still compatible with the observation within the uncertainties of the measured spectrum, specifically the level of the assumed stellar continuum. For derived quantities as the stellar luminosity an error propagation was performed. A more automated analysis is hampered by the many nebular lines which are blended with the stellar lines and the relatively low signal-to-noise (S/N) of about 10 in the blue part of the optical observation.

The stellar temperature $T_\star$, defined at $\tau_\mathrm{Ross,\,cont}=20$, is constrained by the relative strength of the emission lines, mainly by the \ovi\ to \ov\ emission lines ratio. The best fit is achieved with $T_\star=112\pm2$\,kK. For the transformed radius a value of $R_\text{t}=9.1\pm0.7\,R_\odot$ gives the best fit to the optical emission lines.
From the width of the optical emission lines and the O\,{\scshape v}\,1371\,\AA{} P Cygni line in the UV spectrum, we infer a terminal velocity of about 2000$\pm$200~km~s$^{-1}$ using a $\beta$-law with $\beta=1$. Other values of $\beta$ do not improve the fit quality. 
As models with a homogeneous wind give too strong electron-scattering line wings, we use a density contrast of $D=10$, where higher values cannot be excluded \citep[see][]{todt2008}. The resultant mass-loss rate is then $\dot{M}$=(1.6$\pm$0.9)$\times10^{-7}$~M$_\odot$~yr$^{-1}$.

A carbon and helium abundance of $X_\text{He}$:$X_\text{C}$=60:30 by mass is estimated from the almost equal strengths of the He\,{\scshape ii}\,5411\,\AA{} emission line and the C\,{\scshape iv}\,5471\,\AA{} feature. We see however an absorption feature between these two emission lines that cannot be reproduced by our models. Therefore we give rather conservative uncertainties for these abundances: $X_\text{C}=0.30\pm0.10$ and $X_\text{He}=0.60\pm10$ by mass. 
The O\,{\scshape v} and O\,{\scshape vi} lines in the optical spectrum can be best fitted with an abundance of $X_\text{O}=0.15\pm0.05$ by mass. 
We do not see strong spectral lines from the iron group elements in the models in the given parameter range and with regard to the limited quality of the UV spectra, we could not determine an iron abundance and adopted instead the solar value.

We determine a stellar luminosity of $L_\star \approx 12\,000\,L_\odot$ and a reddening of $E_{B-V}\approx 0.8\,$mag from fitting the synthetic SED to the observed UV spectra and optical and IR photometry (see Fig.~\ref{fig:SED_POWR}), using the geometric distance from \cite{BJ2021}. Dust extinction is taken into account by the reddening law of \cite{Cardelli1989} with $R_\text{V}=3.1$. Interstellar Lyman $\alpha$
absorption is calculated with the formalism and the relation between $N_\text{H}$ and $E_{B-V}$ from  \cite{groenlam1989}. 

Finally, a comparison between our {\sc powr} synthetic spectrum and that of a black body with the same effective temperature is presented in Fig.~\ref{fig:SED}. This figure demonstrates that most high energetic photons are absorbed in the stellar atmosphere.

\section{Nebular properties of NGC\,1501}
\label{sec:nebular}

Before attempting a photoionized model of NGC\,1501 we need to fully characterize this WRPN with the most updated methodology. For this, we extracted optical spectra from the WHT ISIS data with the aim to estimate line ratios and its physical properties (electron density $n_\mathrm{e}$ and temperature $T_\mathrm{e}$) and abundances. For comparison, we extracted a spectrum from the same region as that defined in \citet{Ercolano2004}. All line fluxes were measured using Gaussian fitting of the splot task in {\sc iraf} \citep{Tody1993}.

The observed ($F$) and de-reddened ($I$) line fluxes are presented in Table~\ref{tab:obs_lines}. The errors were determined adopting the 1-$\sigma$ deviation, $\sigma_\mathrm{l}$, on each measured flux for each emission line using the formalism described in \citet{Tresse1999}. That is, 
\begin{equation}
\sigma_\mathrm{l} = \sigma_\mathrm{c} D \sqrt{2 N_\mathrm{pix} + \frac{EW}{D}},
\end{equation}
\noindent where $D$ is the WHT ISIS spectral dispersion ($0.439$~\AA~pix$^{-1}$ for the blue arm and $0.829$~\AA~pix$^{-1}$ for the red arm), $\sigma_\mathrm{c}$ is the mean standard deviation per pixel of the continuum around the emission line, $N_\mathrm{pix}$ is the number of pixels covered by the feature and $EW$ is the equivalent width of the line.

The emission lines were identified and measured alongside their uncertainties, which include the statistical and calibration errors. We note that we did not consider in our subsequent analysis the lines showing errors greater than 50 per cent. The plasma analysis, realized to infer the logarithmic extinction ($c$(H$\beta$), electronic temperature and density ($T_\mathrm{e}$, $n_\mathrm{e}$) as well as the ionic and total abundances, was performed with the {\sc PyNeb} routines \citep{Luridiana2015} v.1.1.15b2 with its associated atomic data. Similarly to \citet{Sabin2022}, we adopted a Monte Carlo (MC) scheme with 5000 iterations to determine the uncertainty propagation from the emission lines into the subsequent determination of the physical properties and abundances. With the MC method, the line intensities are taken from a Gaussian distribution around the observed uncorrected values and the standard deviation corresponds in this case to the uncertainty of each line.
All the physical and chemical values presented afterwards correspond to the mean of the distribution and their associated standard deviation.

The resultant estimated extinction was $c$(H$\beta$)=0.68$\pm$0.01, which was obtained adopting a \citet{Cardelli1989} law with $R_\mathrm{V}$=3.1. $T_\mathrm{e}$ was computed using the [O\,{\sc iii}] and [N\,{\sc ii}] lines and $n_\mathrm{e}$ was estimated from the [O\,{\sc ii}]~$\lambda$3727,3729~\AA\ and [S\,{\sc ii}]~$\lambda$6717,6730~\AA\, doublets. The resultant $T_\mathrm{e}$ and $n_\mathrm{e}$ values are also listed at the bottom of Table~\ref{tab:obs_lines}. All ionic and total abundances calculations were performed adopting $T_\mathrm{e}$([O\,{\sc iii}]) and $n_\mathrm{e}$([S\,{\sc ii}]).

While \citet{Sabin2022} used a Machine Learning approach to calculate the Ionization Correction Factors (ICF) specific to their object of study, we will use a more classical approach which involves the ICFs defined by \citet[][hereinafter KB94]{KB1994} and \citet[][hereinafter DIMS14]{DIMS2014} for the subsequent calculation of the total abundances. 
The resultant ionic and total abundances for NGC\,1501 are listed in Table~\ref{tab:abundances} in comparison with those obtained by \citet{Ercolano2004} using a photo-ionization model.
The differences will be discussed in the next section.

\begin{table}
\begin{center}
\setlength{\columnwidth}{0.1\columnwidth}
\setlength{\tabcolsep}{0.4\tabcolsep}
\caption{Observed ($F$) and de-reddened ($I$) emission line fluxes of NGC\,1501.
All line fluxes are normalized with respect to \hb=100. Predictions from our best {\sc cloudy} model are also listed in the last column.}
\begin{tabular}{lcccc}
\hline
\multicolumn{1}{l}{Line} &
\multicolumn{1}{c}{$\lambda_0$} &
\multicolumn{1}{c}{$F$} &
\multicolumn{1}{c}{$I$} &
\multicolumn{1}{c}{Model} \\
\multicolumn{1}{c}{} &
\multicolumn{1}{c}{$($\AA$)$} &
\multicolumn{3}{c}{} \\
\hline
\oii\    & 3727 &  7.41$\pm$0.78  & 12.22$\pm$1.30 &  25.9 \\
\oii\  & 3729 & 6.24$\pm$0.62 &  10.30$\pm$1.02 &  25.1 \\
\hi\  & 3771 &  2.24$\pm$0.46 &  3.65$\pm$0.75 &  5.17 \\
\hi\ & 3798 &  3.11$\pm$0.31 &  5.02$\pm$0.51 & 5.30 \\
\hi\ & 3835 &  5.11$\pm$0.28 &  8.15$\pm$0.45 & 7.34 \\
\neiii\ & 3869 &  69.12$\pm$0.72 &  108.85$\pm$1.57 &  123.20 \\
\hi\ & 3889 &  12.27$\pm$1.53 &  19.17$\pm$2.40 &  10.59 \\
\neiii\  & 3968 &  20.11$\pm$0.52 &  30.47$\pm$0.84 & 39.10 \\
\hi\ & 3970 &  12.15$\pm$0.09 &  18.39$\pm$0.22 &  16.33 \\
\hei\ & 4026 &  1.70$\pm$0.58 &  2.52$\pm$0.86 &  2.24 \\
\hi\ & 4102 &  21.46$\pm$0.40 &  30.70$\pm$0.63 & 26.37 \\
\heii\ & 4198 &  0.31$\pm$0.10 &  0.42$\pm$0.13 & 0.12 \\
\cii\ & 4267 &  0.82$\pm$0.49 &  1.09$\pm$0.64 & 0.03\\
\hi\ & 4341 &  38.54$\pm$0.65 &  49.18$\pm$0.89 & 47.83 \\
\oiii\   & 4363 &  8.73$\pm$0.46 &  11.02$\pm$0.59 & 15.10 \\
\hei\    & 4388 &  1.18$\pm$0.09 &  1.48$\pm$0.11 & 0.60  \\
\hei\    & 4471 &  3.51$\pm$0.41 & 4.20$\pm$0.49 & 4.54 \\
\heii\ & 4541 &  1.03$\pm$0.17 & 1.19$\pm$0.20 & 0.17 \\
\niii\ & 4641 &  0.84$\pm$0.08 & 0.93$\pm$0.08 & $<$0.01 \\
\ciii\ & 4649 &  0.45$\pm$0.06 & 0.49$\pm$0.07 & $<$0.01 \\
\heii\ & 4686 &  39.68$\pm$0.40 & 42.87$\pm$0.45 & 4.64 \\
\ariv\   & 4712 &  2.42$\pm$0.10 & 2.58$\pm$0.11 & 2.38 \\
\ariv\   & 4741 &  1.99$\pm$0.12 & 2.10$\pm$0.11 & 1.74 \\
\hb\ & 4861 &  100 & 100 & 100\\
\oiii\   & 4959 &  386.42$\pm$2.29 & 371.17$\pm$2.03 & 385.52 \\
\oiii\   & 5007 &  1204.27$\pm$7.09 & 1135.24$\pm$1.40 & 1150.1 \\
\cliii\  & 5518 &  1.52$\pm$0.15 & 1.21$\pm$0.12 & 1.48 \\
\cliii\  & 5538 &  1.38$\pm$0.13 & 1.10$\pm$0.10 & 1.22 \\
\nii\    & 5755 &  0.73$\pm$0.12 & 0.55$\pm$0.09  & 0.43 \\
\civ\  & 5801 &  0.13$\pm$0.01 & 0.10$\pm$0.01 & $<$0.01 \\
\civ\  & 5812 &  0.12$\pm$0.01 & 0.09$\pm$0.01 & $<$0.01 \\
\hei\ & 5876 &  16.23$\pm$0.20 & 11.82$\pm$0.14 & 12.38 \\
\oi\   & 6300 &  0.49$\pm$0.07 & 0.32$\pm$0.05 & $<$0.01 \\
\siii\ & 6312 &  2.85$\pm$0.13 & 1.89$\pm$0.09 & 2.40 \\
\nii\    & 6548 &   5.50$\pm$0.14 & 3.47$\pm$0.09 & 4.50 \\
\ha\     & 6563 &  453.44$\pm$2.71 & 285.00$\pm$0.00 &  282.30 \\
\nii\    & 6584 &  15.60$\pm$0.23 & 9.75$\pm$0.14 & 13.27 \\
\hei\    & 6678 &  6.85$\pm$0.16 & 4.20$\pm$0.09 & 3.4 \\
\sii\    & 6717 &  4.26$\pm$0.14 & 2.59$\pm$0.08 & 3.76 \\
\sii\    & 6731 &  4.40$\pm$0.14 & 2.67$\pm$0.08 & 4.04 \\
\hei\    & 7065 &  5.82$\pm$0.10 &  3.30$\pm$0.05 & 3.50 \\
\ariii\  & 7136 &  27.86$\pm$0.24 & 15.55$\pm$0.11 & 15.93 \\
\heii\ & 7177 &  1.20$\pm$0.08 & 0.67$\pm$0.04 & $<$0.01 \\
\cii\ & 7231 &  0.39$\pm$0.06 & 0.21$\pm$0.03 & $<$0.01 \\
\ariv\ & 7237 &  1.08$\pm$0.06 & 0.59$\pm$0.03 & $<$0.01 \\
\hei\ & 7281 &  1.43$\pm$0.07 & 0.77$\pm$0.04 & 0.74 \\
\oii\ & 7320 &  1.84$\pm$0.08 & 0.99$\pm$0.04 & 0.88 \\
\oii\ & 7330 &  1.90$\pm$0.08 & 1.02$\pm$0.08 & 0.50 \\
\cliv\ & 7531 &  0.82$\pm$0.20 & 0.42$\pm$0.10 & 0.26 \\
\heii\ & 7592 &  1.50$\pm$0.20 & 0.76$\pm$0.10 & 0.08 \\
\ariii\ & 7751 &  8.42$\pm$0.28 & 4.14$\pm$0.03 & 3.78  \\
\hline
\multicolumn{1}{l}{log($F$(H$\beta$))} & \multicolumn{1}{l}{[erg~cm$^{-2}$~s$^{-1}$] } &
\multicolumn{2}{c}{$-$10.54$\pm$0.01} & \multicolumn{1}{c}{$-$10.33} \\
\multicolumn{1}{l}{$c$(H$\beta$)} & \multicolumn{1}{l}{ } &
\multicolumn{2}{c}{0.68$\pm$0.01} & \multicolumn{1}{c}{0.68} \\
\hline
$T_\mathrm{e}$(\oiii)  & [K]         &  11390$\pm$210 & & 12600 \\
$T_\mathrm{e}$(\nii)   & [K]         &  18200$\pm$2200 & &  14370 \\
$n_\mathrm{e}$ (\sii)  & [cm$^{-3}$] &  840$\pm$150 & & 940 \\
$n_\mathrm{e}$ (\ariv) & [cm$^{-3}$] & 1300$\pm$600 & & 260 \\
$n_\mathrm{e}$ (\cliii)& [cm$^{-3}$] & 1600$\pm$900 & &  890\\
\hline
\end{tabular}
\label{tab:obs_lines}
\vspace{0.15cm}
\end{center}
\end{table}

\begin{table}
\begin{center}
\caption{Ionic and total abundances of NGC\,1501. Total abundances are expressed as 12 + log(X/H)}
\setlength{\tabcolsep}{0.8\tabcolsep}    
\begin{tabular}{lccc}
\hline
Ionic abundances  \\
\hline
\multicolumn{1}{l}{Ar$^{+2}$/H$^{+}$ } &
\multicolumn{3}{c}{ (4.39$\pm$0.09)$\times10^{-7}$} \\
\multicolumn{1}{l}{Ar$^{+3}$/H$^{+}$} &
\multicolumn{3}{c}{ (2.66$\pm$0.27)$\times10^{-6}$} \\
\multicolumn{1}{l}{Cl$^{+2}$/H$^{+}$} &
\multicolumn{3}{c}{ (3.61$\pm$0.95)$\times10^{-8}$} \\
\multicolumn{1}{l}{Cl$^{+3}$/H$^{+}$} &
\multicolumn{3}{c}{ (8.10$\pm$1.99)$\times10^{-8}$} \\
\multicolumn{1}{l}{He$^{+}$/H$^{+}$} &
\multicolumn{3}{c}{ (9.92$\pm$0.73)$\times10^{-2}$} \\
\multicolumn{1}{l}{He$^{+2}$/H$^{+}$} &
\multicolumn{3}{c}{ (3.57$\pm$0.20)$\times10^{-2}$} \\
\multicolumn{1}{l}{N$^{+}$/H$^{+}$} &
\multicolumn{3}{c}{ (7.25$\pm$1.63)$\times10^{-7}$} \\
\multicolumn{1}{l}{Ne$^{+2}$/H$^{+}$} &
\multicolumn{3}{c}{ (6.91$\pm$0.47)$\times10^{-5}$} \\
\multicolumn{1}{l}{O$^{0}$/H$^{+}$} &
\multicolumn{3}{c}{ (1.02$\pm$0.39)$\times10^{-7}$} \\
\multicolumn{1}{l}{ O$^{+}$/H$^{+}$} &
\multicolumn{3}{c}{ (3.14$\pm$1.72)$\times10^{-6}$} \\
\multicolumn{1}{l}{O$^{+2}$/H$^{+}$} &
\multicolumn{3}{c}{ (2.57$\pm$0.14)$\times10^{-4}$} \\
\multicolumn{1}{l}{S$^{+}$/H$^{+}$} &
\multicolumn{3}{c}{ (4.77$\pm$0.97)$\times10^{-8}$} \\
\multicolumn{1}{l}{S$^{+2}$/H$^{+}$} &
\multicolumn{3}{c}{( 6.03$\pm$2.21)$\times10^{-7}$} \\
\hline
Total abundances \\
Element & This work &  \citet{Ercolano2004} & Model \\
\hline
He &  11.13$\pm$0.02 & 11.01  & 11.01 \\
C  &                 &  8.53  & 8.39 \\ 
O  &  8.49$\pm$0.03  &  8.53  & 8.20 \\
N  &  7.91$\pm$0.10  &  8.52  & 7.60 \\
Ne &  7.91$\pm$0.03  & 7.84   & 7.88 \\
Cl & 5.07$\pm$0.06   & 5.10   & 5.07 \\
S  & 6.88$\pm$0.12   & 7.52   & 6.50 \\
Ar & 5.91$\pm$0.09   & 6.11   & 6.04 \\
\hline
\end{tabular}
\label{tab:abundances}
\end{center}
\end{table}

\section{Photoionization model of NGC\,1501}
\label{sec:photoionization}

To produce an updated view of the nebular and dust properties of NGC\,1501 we used the photoionization code {\sc cloudy} \citep{Ferland2017}. {\sc cloudy} requires as inputs: i) the ionization source (the stellar atmosphere of the central star), ii) the gas distribution and characteristics (geometry, density and chemical abundances) and iii) the dust properties (species, spatial distribution and grain sizes). In the following we address the different properties that resulted in the best model for the dust and gas in NGC\,1501.

Now that we have a specific stellar atmosphere model for WD\,0402+607 (see Sec.~\ref{sec:PoWR}) and the physical properties of the gas in NGC\,1501 (see Sec.~\ref{sec:nebular}), we can use {\sc cloudy} to produce the most accurate model of this WRPN including gas and dust. Synthetic optical and IR observations can be simulated making use of the {\sc pyCloudy} routines \citep{Morisset2013}. {\sc pyCloudy} allow us to create synthetic long-slit observations with the same aperture and position as those shown in Figure~\ref{fig:HST}, which allow us to compare directly with values obtained with the WHT ISIS observations.

\begin{figure}
\begin{center}
  \includegraphics[width=0.8\linewidth]{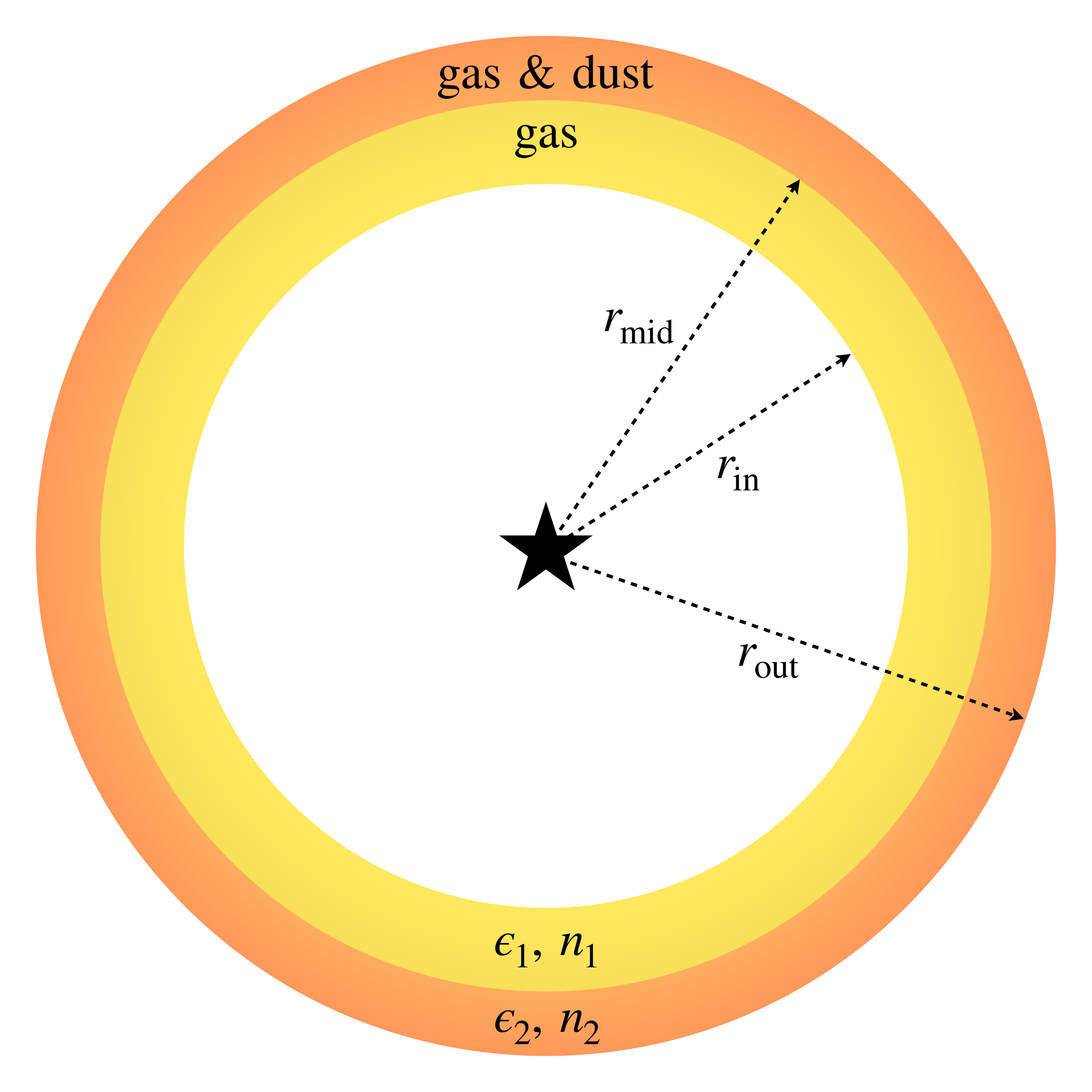}
\caption{Schematic view of the two-shell geometry used to model the nebular and dust properties of NGC\,1501. The inner shell 
accounts for the presence of gas while the outer shell contains gas and dust.}
\label{fig:esquema}
\end{center}
\end{figure}

We started our modelling by defining a spherical shell distribution of material. We defined an outer radius of $r_{\mathrm{out}}=30^{\prime\prime}$ which is an averaged value for NGC\,1501. C-rich dust was taken into account to be present in the nebula. In particular, we included in our calculation amorphous carbon which is precompiled in the {\sc cloudy} code. A large number of models were attempted in order to assess the inner radius ($r_\mathrm{in}$), but we note that the model improved after we defined a double-shell structure with gas and dust in the outer shell and only gas in the inner shell. This scheme is illustrated in Figure~\ref{fig:esquema}.

Our best {\sc cloudy} model of NGC\,1501 consists of two spherical shells with external radius $r_{\mathrm{out}}=30^{\prime\prime}$, inner radius $r_{\rm{in}}=20^{\prime\prime}$ and mid radius separating the two shells $r_{\rm{mid}}=26^{\prime\prime}$. The inner shell contains gas with a constant density profile with $n_1=400$ cm$^{-3}$ and a filling factor of $\epsilon_1=0.05$. For the outer shell, a constant density profile $n_2=650$~cm$^{-3}$ with a filling factor of $\epsilon_2$= 0.5 is sufficient to reproduce the nebula's optical emission. Synthetic intensity ratios provided by the model are listed in the last column of Table~\ref{tab:obs_lines} for its comparison with those of the emission lines detected in the spectrum. The values of $T_\mathrm{e}$ and $n_\mathrm{e}$ derived from diagnostic-sensitive line ratios of the model are also presented at the bottom of Table~\ref{tab:obs_lines}.

It is worth noticing a few discrepancies between our model predictions and observed emission lines. First, we note that our {\sc Cloudy} model underpredicts the He\,{\sc ii} lines. Form example, the He\,{\sc ii} 4686~\AA\, is underpredicted by an order of magnitude. As illustrated by Fig.~\ref{fig:SED} most energetic photons are absorbed by the dense winds of WR-type stars. Thus, we suggest that the detected emission from He\,{\sc ii} lines might be produced by shocks, very likely at the inner region of the nebular shell where is pressure-driven by the hot bubble. On the other hand, the [O\,{\sc ii}] lines are overpredicted by a factor of $\sim$2. This problem could be alleviated by adopting a multi-layer density distribution or a model including dense clumps embedded within different density structures, but such a complex model is out of the scope of the present work.

The outer shell has a mixture of amorphous carbonaceous dust with a power-law size distribution $\propto a^{-3.5}$ \citep{Mathis1977} with ten size bins. The observed SED is successfully reproduced by including two population of dust sizes as shown in Fig.~\ref{fig:sed_cont}. A distribution of small grains with sizes $a_\mathrm{small}$=[0.001--0.002]~$\mu$m and another with sizes $a_\mathrm{large}$=[0.03--0.08]~$\mu$m are needed to achieve the emission at $\sim$25 and $\sim$90~$\mu$m respectively. Through an iterative process we found that a 5:1 large-to-small dust ratio is needed to reproduce the IR and radio emission. The synthetic SED obtained with our best {\sc cloudy} model is compared with the observed IR and radio SED in Fig.~\ref{fig:sed_cont}.

A comparison of the total abundances for NGC\,1501 derived using {\sc PyNeb} in the previous section and those derived from photo-ionization models here and by \citet{Ercolano2004} is provided in Table~\ref{tab:abundances}.  
In general, the total abundances derived from the two photo-ionization models agree, except for nitrogen and sulfur, which are much higher in the model computed by \citet{Ercolano2004} that implies ICFs about 11 times larger than the ones derived from our model.
Otherwise, the difference in the helium abundances derived by {\sc PyNeb} and by photo-ionization models can most likely be attributed to the adoption of a higher electron temperature to compute the He$^+$ abundances by {\sc PyNeb}. Finally, we would like to notice that we are not able to compute C abundances. 
The value of 8.53 estimated by \citet{Ercolano2004} is derived from recombination lines, which are strongly enhanced in NGC\,1501 and probably not representative of the bulk nebular abundances. \citet{Ercolano2004} assumed the presence of unresolved H-deficient clumps in NGC\,1501 to try to explain such high abundance discrepancy factors. To avoid any strong assumptions, we set the C abundance to the solar value of 8.39 \citep[see][]{Lodders2009}.

\subsection{Photoionization model implications}

Our {\sc cloudy} model of NGC\,1501 makes a good description of the optical, IR and radio observables of this WRPN. The model is able to reproduce most of the optical emission lines detected in the WHT ISIS spectra as well as the publicly available IR and radio measurements, with differences within the errors (see Fig.~\ref{fig:sed_cont} and Table~\ref{tab:obs_lines}). The resultant physical properties predicted by the model ($n_\mathrm{e}$ and $T_\mathrm{e}$) are also within those estimated from observations. 

\begin{table}
\begin{center}
\caption{{\scshape Cloudy} model results.}
\begin{tabular}{lcc}
\hline
\multicolumn{1}{c}{Parameter} &
\multicolumn{2}{c}{Model} \\
\hline
\multicolumn{1}{c}{$M_{\mathrm{TOT}}$ [M$_{\odot}$] } &
\multicolumn{2}{c}{2.182$\times10^{-1}$} \\
\multicolumn{1}{c}{ $M_{\mathrm{Gas}}$ [M$_{\odot}$] } &
\multicolumn{2}{c}{ 2.116$\times10^{-1}$} \\
\multicolumn{1}{c}{$M_{\mathrm{Dust}}$ [M$_{\odot}$] } &
\multicolumn{2}{c}{ 8.908$\times10^{-4}$ } \\
\hline
\multicolumn{1}{c}{} & \multicolumn{1}{c}{Size distribution} & \multicolumn{1}{c}{Mass [M$_{\odot}$]} \\
\multicolumn{1}{c}{$a_\mathrm{large}$}   & \multicolumn{1}{c}{0.030--0.080 $\mu$m } & \multicolumn{1}{c}{ 7.423$\times10^{-4}$} \\
\multicolumn{1}{c}{$a_\mathrm{small}$} & \multicolumn{1}{c}{0.001--0.002 $\mu$m}  &  \multicolumn{1}{c}{ 1.484$\times10^{-4}$} \\
\hline
\end{tabular}
\label{tab:model_results}
\end{center}
\end{table}

Our {\sc cloudy} model predicts total ionized and dust masses for NGC\,1501 of $\sim$0.22~M$_{\odot}$ and 8.9$\times$10$^{-4}$~M$_\odot$, respectively, which corresponds to a gas-to-dust ratio of $\approx$240.  
This ionized mass is similar to that derived by \citet{Santander2022}, 0.2$^{+0.20}_{-0.11}$~M$_\odot$, who also provided an upper limit $<$0.2~M$_\odot$ for the molecular content of NGC\,1501. The specific mass contributions obtained from our {\sc cloudy} model are listed in Table~\ref{tab:model_results}. 
Adopting a current stellar mass for WD\,0402+607 of 0.60$^{+0.06}_{-0.02}$~M$_\odot$ as that estimated by \citet{Corsico2021}, a lower limit to the mass of the progenitor can be estimated to be in the 0.80--0.88~M$_{\odot}$ range by accounting for the gas and dust masses.

Our estimation of the mass of the progenitor of NGC\,1501 is relatively low if one takes into account that PNe can form from the evolution of stars as massive as 8~M$_\odot$. This apparent discrepancy is a known problem and has been described as {\it the missing mass problem} \citep{Kwok1994}, which has been addressed with several ideas. Some authors have argue that low-mass stars with high proper motions might leave behind the mass expelled during their evolution \citep[see, e.g.,][and references therein]{Villaver2012}, whilst other have suggested that a fraction of the mass is in the form of molecular material \citep[e.g.,][]{BST2005,Kimura2012}. However, neither of these ideas seems to apply to NGC\,1501. No asymmetries produced by a possible strong proper motion \citep{Gurzadian1969} is appreciated in this WRPN. Furthermore, there are no determinations of the molecular content of NGC\,1501 in the literature, only an upper limit of 0.2~M$_\odot$ has been estimated \citep{Santander2022}. Nevertheless, the abundances of NGC\,1501 might also help peering into the mass of its progenitor.

\begin{figure*}
\begin{center}
  \includegraphics[width=\linewidth]{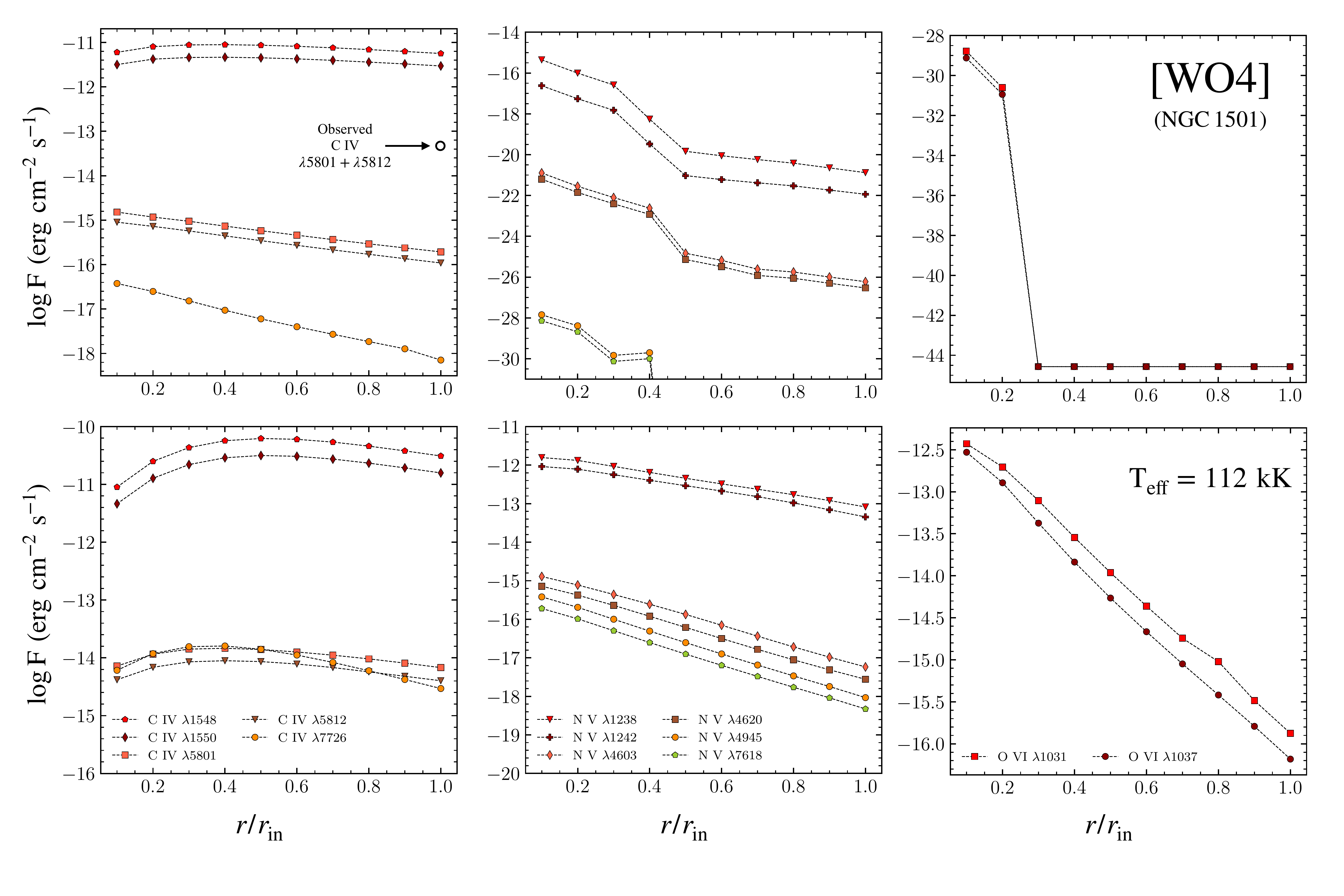}
\caption{Comparison between synthetic emission lines fluxes of different ions obtained with {\sc cloudy} for the best fit to the CSPN of NGC\,1501 (top panels) and a black-body with T$_{\rm{eff}}=$112 kK (bottom panels). The left, middle and right panels show the predictions from different C\,{\sc iv}, N\,{\sc v} and O\,{\sc vi} emission lines, respectively. Different symbols represent different emission lines. The radius is normalized to $r_\mathrm{in}$=26$''$(=0.22~pc), so that $r/r_\mathrm{in}$=1 corresponds to the best {\sc cloudy} model of NGC\,1501. All models were run with a fixed shell thickness of $\Delta r=4''$($\approx$0.03~pc). For comparison the upper-left panel shows the current observed fluxes from C\,{\sc iv} at 5801 and 5812 \AA.}
\label{fig:lineas_cloudy}
\end{center}
\end{figure*}

\citet{GR2013} calculated the abundances of a set of WRPNe and found an empiric relation between the N and O abundances. These authors found a relation
\begin{equation}
\mathrm{log(N/O)} = 0.73 \times [12+\mathrm{log(N/H)}] - 6.5,   
\end{equation}
\noindent for the N and O abundances that can be interpreted in terms of the stellar evolution models from \citet{Karakas2010}. Accordingly, the N and O abundances of NGC\,1501 imply an initial mass for its progenitor in the low-mass range of $\lesssim$1~M$_\odot$ \citep[see fig.~6 in][]{GR2013}. 

We then propose that in fact the progenitor mass of NGC\,1501 was $\lesssim$1~M$_\odot$ which is suggested independently by our photoionization model and the abundance determinations. We note that a similar result was proposed for the WRPN NGC\,6905 by \citet{GG2022}.

The reprocessed WHT ISIS spectrum allowed us to estimate errors in emission lines intensities that had previously not been reported. 
These errors are used to adopt weights and compute error bars in the estimate of ionic chemical abundances, ICFs, and total abundances. 
The abundances of O, Ne, Cl and Ar presented in this work are consistent with those previously reported by \citet{Ercolano2004}, whereas those of N  and S are an order of magnitude lower. These differences can be attributed to variations of the values of their ICFs caused by different values of the ionic abundances of O$^+$, N$^+$, and S$^+$ used to compute them, which can be explained by the smaller extinction value and different line intensity weights and atomic parameters \citep[see][and references therein]{Sabin2022} used in this work.

\section{The mixing layer in NGC\,1501}
\label{sec:mix_ngc1501}

\begin{figure}
\begin{center}
  \includegraphics[width=\linewidth]{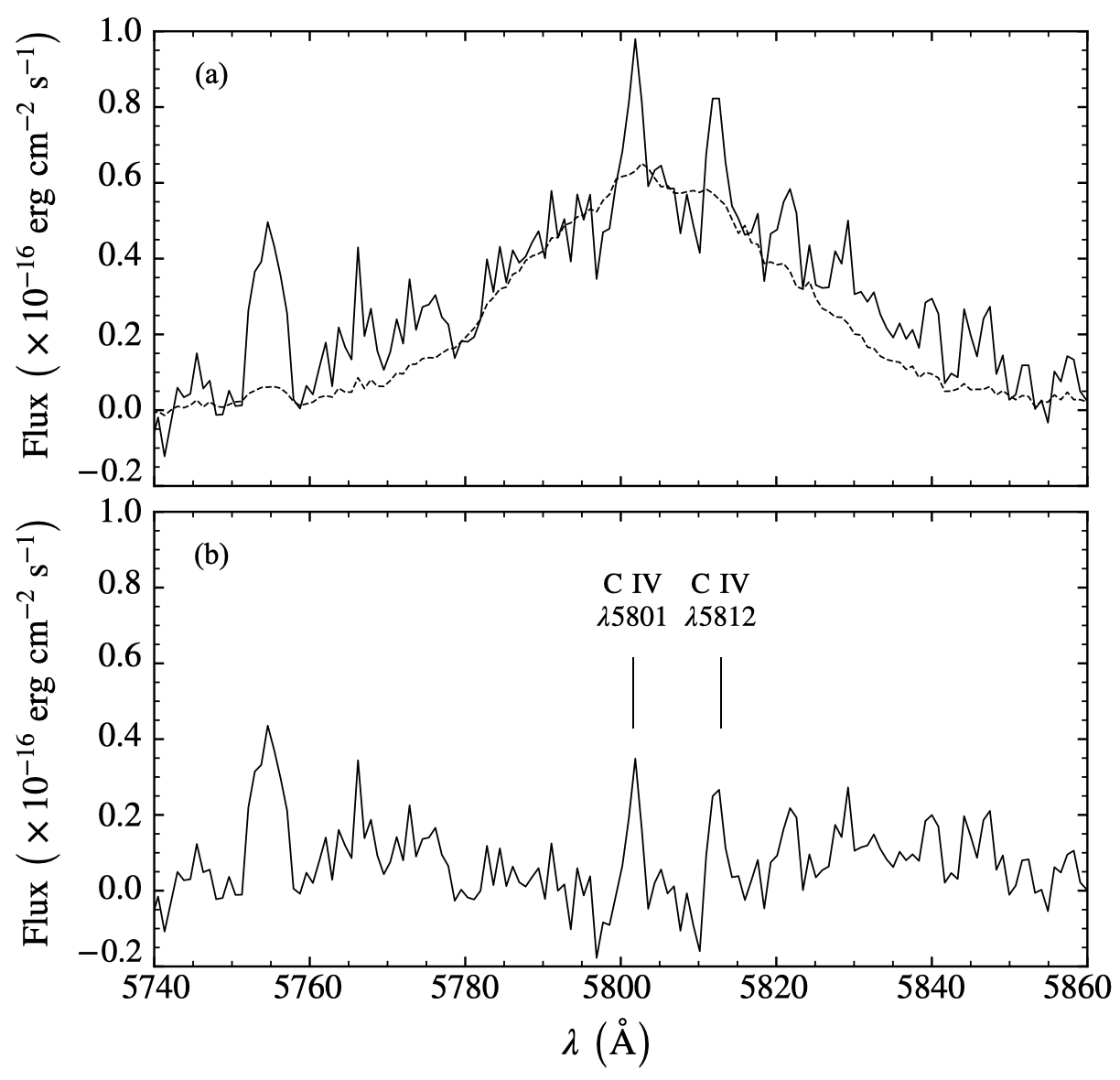}
\caption{(a) Solid line corresponds to the nebular spectrum  while the dashed line shows the broad  C\,{\sc iv} dispersed component in the nebular spectrum.\\
(b) Nebular emission after subtracting the diffuse stellar contribution.}
\label{fig:linea_civ}
\end{center}
\end{figure}

The N\,{\sc v} emission lines at 4603, 4620 and 4945~\AA\, are detected at the background level in the WHT ISIS spectra, whilst N\,{\sc v} emission line at 7618~\AA\ is contaminated by the telluric Fraunhofer A-band caused by molecular O$_2$ absorption therefore any detection at this spectral range without appropriate telluric correction seems to be spurious.
However, the C\,{\sc iv} emission lines at 5801,5812~\AA\ are detected as narrow nebular features. This emission, however, is contaminated by the broad C\,{\sc iv} stellar emission dispersed by the spectrograph (see Fig.~\ref{fig:linea_civ} top panel). To subtract this contamination from the nebular spectrum, we have scaled the broad C\,{\sc iv} stellar component to the broad C\,{\sc iv} dispersed component in the nebular spectrum and subtracted it to obtain a net nebular spectrum in this spectral range. This is illustrated in Fig.~\ref{fig:linea_civ} bottom panel.
We note that a direct comparison with estimates from \citet{Ercolano2004} is impossible because these authors did not reported errors in their measurements.

Our {\sc cloudy} model is not able to reproduce the C\,{\sc iv} 5801,5812 nebular emission lines reported in the WHT ISIS spectrum of NGC\,1501. According to the photoionization model, they should be largely below the background level. That is, their origin is not due to photoionization and we attribute them to the presence of a mixing layer.

To assess the possible contribution from the mixing layer, we extended our {\sc cloudy} calculations to estimate the emission from C\,{\sc iv}, N\,{\sc v} and O\,{\sc vi} in UV wavelengths. We used our stellar atmosphere model of WD\,0402+607 obtained with {\sc powr} to run a family of models. We started with the same geometry used in our best {\sc cloudy} model, but varied the inner radius from 2.6$''$(=0.022~pc) to 26$''$(=0.22~pc), while keeping the thickness of the shell fixed to $\Delta r$=4$''$($\approx$0.03~pc). Fig.~\ref{fig:lineas_cloudy} top row panels show the spatial distribution of several C\,{\sc iv}, N\,{\sc v} and O\,{\sc vi} in the optical and UV wavelengths. The radial axis is presented in units of $r/r_\mathrm{in}$ so that a value of 1 represents the best {\sc cloudy} model obtained for NGC\,1501. For comparison, the bottom panels of this figure also shows the results obtained by adopting a blackbody model with the same temperature as estimated for WD\,0402+607 ($T=$112~kK) instead of the detailed atmosphere model (see Fig.~\ref{fig:SED} and Sec.~\ref{sec:PoWR}). 

Fig.~\ref{fig:lineas_cloudy} shows that the dominant lines should be those in the UV in accordance with the atomic transition probabilities of these elements \citep[e.g.,][]{Wiese1996}. Specifically, the C\,{\sc iv}$\lambda\lambda$1548,1550~\AA\, lines are slightly more intense for the blackbody model than the detailed atmosphere model. The difference is more evident for the optical C\,{\sc iv} 5801, 5812 and 7726~\AA\, lines. These have fluxes ranging  $\lesssim10^{-14}$~erg~s$^{-1}$~cm$^{-2}$ for the blackbody model, at least one order of magnitude more intense when compared with results from the detailed stellar atmosphere model in the top row of Fig.~\ref{fig:lineas_cloudy}. The situation is more dramatic when comparing the fluxes from the N\,{\sc v} and O\,{\sc vi} lines in the middle and right panels of Fig.~\ref{fig:lineas_cloudy}. The lines are predicted to be orders of magnitude more intense for the case of the blackbody model, which reaffirms the need to include detailed stellar atmosphere models when producing photoionization studies of WRPNe.

A comparison with the UV observations of NGC\,1501 shows that our photoionization model is not enough to even produce the C\,{\sc iv}$\lambda\lambda$1548,1551~\AA\, doublet. Our model predicts a combined total flux for these two lines below 10$^{-11}$~erg~s$^{-1}$~cm$^{-2}$, whereas the value estimated for the unresolved doublet is 3.2$\times10^{-11}$~erg~s$^{-1}$~cm$^{-2}$ \citep{feibelman1998b}. In addition, the observed fluxes of the C\,{\sc iv} 5801, 5812~\AA\, emission lines are an order of magnitude more intense than those predicted by our best {\sc cloudy} model as shown with the empty circle mark in the upper-left panel in Fig.~\ref{fig:lineas_cloudy}. This result reinforces the idea of the presence of a mixing layer in NGC\,1501 enhancing the emission from C\,{\sc iv} lines in the UV and optical wavelengths. Unfortunately, there are no detection of the N\,{\sc v} lines in the UV to make similar assessments, and the optical N\,{\sc v} lines are detected at the noise level.

The intensity of the C\,{\sc iv} 5801$\lambda$ line can be estimated as \citep[see][]{Fang2016}
\begin{equation}
    I=n_{\rm{e}}n_{\rm{C}^{3+}}h\nu\frac{8.629\times10^{-6}}{\sqrt{T_\mathrm{e}}} \frac{\Omega(1,2;T_\mathrm{e})}{g_1} e^{-\chi/k_\mathrm{B}T_\mathrm{e}}\frac{V}{4\pi d^2},
\end{equation}
\noindent 
where $\chi$ is the excitation energy of the upper level ($\chi$=39.68 eV), $V$ the volume of the emitting shell, $n_\mathrm{e}$ and $n_\mathrm{C^{3+}}$ are the number densities of the electron and the C$^{3+}$ ion respectively, $\Omega(1,2;T_\mathrm{e})$ is the Maxwellian-averaged collision strength of C$^{3+}$ 3$s^{2}S_{1/2}$--3$p^{2}P^{0}_{3/2}$, $g_1$ is the statistical weight for the lower level (for C\,{\sc iv} 5801$\lambda$, $g_1$=2). 
The largest ionization fraction $\approx$30\% of C$^{+3}$ is reached at an optimal electron temperature of 100,000 K \citep[CHIANTI v.10 data;][]{delzanna2021}. 
Thus, adopting the carbon abundance of 12+log(C/H)=8.53 provided by \citet{Ercolano2004}, a $n_{\rm{C}^{3+}}$=1.0$\times$10$^{-4}$ can be derived.

Assuming that the UV emitting regions is spatially coincident with the inner shell defined by our {\sc cloudy} model, that is, with inner and outer radii of 20 and 26$''$, this corresponds to a volume of 5.9$\times10^{53}$~cm$^{-3}$. Introducing a filling factor $\epsilon$, the electron density as a function of temperature is $n_\mathrm{e}=0.11\ T_\mathrm{e}^{1/4}\epsilon^{-1/2}\exp{(2.3\times 10^{5}/T_\mathrm{e})}$. 
The density in the interface layer is close to $\sim11.2\epsilon^{-1/2}$ cm$^{-3}$ for the adopted temperature of $10^{5}$ K.
Thus, the pressure of the mixing region can be estimated to be $P_\mathrm{mix}\approx2\times10^{-10}\epsilon^{-1/2}$~dyn~cm$^{-2}$.

On the other hand, we used the available {\it Chandra} observations \citep[see][]{Freeman2014} to extract and analyse the spectrum of the hot bubble in NGC\,1501. The best model required a plasma temperature of $T_\mathrm{X}=$1.8$\times10^{6}$~K and resulted in $n_\mathrm{e,X}=2.7\epsilon^{-1/2}$~cm$^{-3}$. That is, the hot gas has a pressure of $P_\mathrm{X}=6.5\times10^{-10}\epsilon^{-1/2}$~dyn~cm$^{-2}$, very similar to that of the mixing layer. This means that the mixing layer is in pressure equilibrium with the hot bubble confirming that they share the same physics. It is very likely that the small differences within the pressure values are given by the clumpy distribution of each emitting region.

\begin{table*}
\begin{center}
\caption{Logarithm of the emission-line fluxes of different WRPNe taken from UV spectra from the literature. The spectra have been corrected for extinction. The distances were taken from \citet{BJ2021}.}
\setlength{\tabcolsep}{0.8\tabcolsep}    
\begin{tabular}{lccccccccccc}
\hline
\multicolumn{1}{l}{Object} &
\multicolumn{1}{c}{[WR]-type} &
\multicolumn{1}{c}{$d$} & $r$ &
\multicolumn{2}{c}{\nv} &
\multicolumn{2}{c}{\civ} &
\multicolumn{1}{c}{$c$(H$\beta$)} & \multicolumn{1}{c}{X-rays} & \multicolumn{1}{c}{Mixing} & \multicolumn{1}{c}{Ref.} \\
\multicolumn{1}{l}{} &
\multicolumn{1}{c}{} &
\multicolumn{1}{c}{(kpc)} & ($''$) & 
\multicolumn{1}{c}{1238~\AA} &
\multicolumn{1}{c}{1241~\AA} &
\multicolumn{1}{c}{1548~\AA} &                    
\multicolumn{1}{c}{1551~\AA} &
\multicolumn{1}{c}{} & \multicolumn{1}{c}{} & \multicolumn{1}{c}{Reg.} & \multicolumn{1}{c}{} \\
(1) & (2) & (3) & (4) & (5) & (6) & (7) & (8) & (9) & (10) & (11) & (12) \\
\hline
NGC\,2452 & \multicolumn{1}{c}{[WO1]} & \multicolumn{1}{c}{4.54} & 10 & \multicolumn{2}{c}{$-$11.50} & \multicolumn{2}{c}{$-$10.40} & \multicolumn{1}{c}{0.77} & \xmark & \cmark & \multicolumn{1}{c}{1, 2} \\
NGC\,5189 & [WO1]  & 1.40 & 75 & ... & .... & $-$11.18 & $-$11.44 & 0.47 & \cmark & \cmark & \multicolumn{1}{c}{3, 4, 5} \\
NGC\,2867 & [WO2]  & 2.85 & 7 & ... & .... & $-$10.37 & $-$10.76 & 0.26 & \xmark & \xmark & \multicolumn{1}{c}{6, 7}\\
NGC\,6905 & [WO2]  & 2.66 & 35 & \multicolumn{2}{c}{$-$12.27} & \multicolumn{2}{c}{$-$10.63} & \multicolumn{1}{c}{0.15} & \xmark & \cmark & \multicolumn{1}{c}{8, 9} \\
Hen 2-55 & \multicolumn{1}{c}{[WO3]} & 8.72 & 10 & \multicolumn{2}{c}{$-$11.85} & \multicolumn{2}{c}{$-$12.91} & \multicolumn{1}{c}{} & \xmark & ? & \multicolumn{1}{c}{10} \\
NGC\,1501 & \multicolumn{1}{c}{[WO4]} & 1.66 & 20 & \multicolumn{2}{c}{....} & \multicolumn{2}{c}{$-$10.49} & \multicolumn{1}{c}{0.68} & \cmark & \cmark & \multicolumn{1}{c}{This work, 11, 12} \\
NGC\,5315 & \multicolumn{1}{c}{[WO4]} & 1.40 & 1 & \multicolumn{2}{c}{$-$10.74} & \multicolumn{2}{c}{$-$9.55} & \multicolumn{1}{c}{0.60} & \cmark & \cmark & \multicolumn{1}{c}{13, 14, 15} \\
\multicolumn{1}{l}{NGC\,6751} & \multicolumn{1}{c}{[WO4]} & 3.09 & 12 & \multicolumn{2}{c}{...} & \multicolumn{2}{c}{$-$9.07} & \multicolumn{1}{c}{0.87} & \xmark & \cmark &
\multicolumn{1}{c}{16, 17} \\
\hline
\end{tabular}
\label{tab:uv_fluxes}
\end{center}
References: (1) \citet{kaler1976}, (2) \citet{feibelman1999}, (3) \citet{GR2012}, (4) \citet{Toala2019a}, (5) \citet{feibelman1997}, (6) \citet{keller2014}, (7) \citet{feibelman1998a}, (8) \citet{GG2022}, (9) \citet{feibelman1996}, (10) \citet{feibelman1995a}, (11) \citet{feibelman1998b}, (12) \citet{Freeman2014}, (13) \citet{feibelman1998c}, (14) \citet{feibelman1998c}, (15) \citet{Kastner2008}, (16) \citet{feibelman1995b}, and (17) \citet{koesterke1997}.\\
NOTES: Column (10) denote those WRPNe that have been observed and detected by X-ray telescopes, whilst (11) shows our predictions for the presence of a mixing layer.
\end{table*}

\begin{figure*}
\begin{center}
  \includegraphics[width=\linewidth]{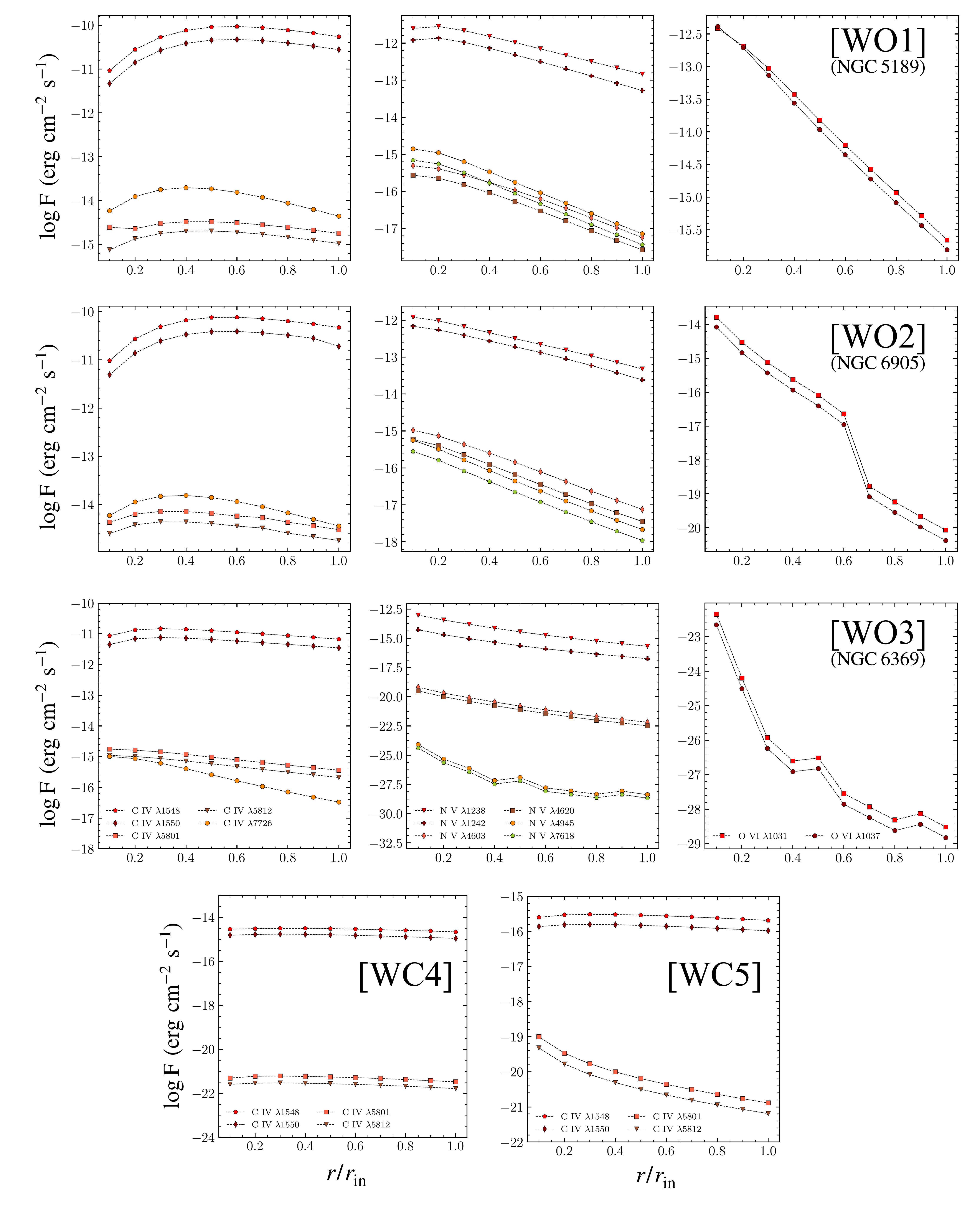}
\caption{Comparison between {\sc cloudy} synthetic emission lines fluxes of different ions from [WR]-type {\sc powr} stellar atmosphere models. Left panel:C IV 1548, 1550 , 5801, 5812 and 7726 \AA. Center panel: N V 1238, 1242, 4603, 4620, 4945  and 7618 \AA. Right panel: O VI 1031 and 1037 \AA. For each model we decrease the internal radius ($r_{\mathrm{in}}$) of the shell. All models has the same  thickness ($\Delta r=4''$).}
\label{fig:lineas_cloudy2}
\end{center}
\end{figure*}

Thus far, there is only one other PN for which a detailed characterization of its mixing layer has been presented in the literature, NGC\,6543 \citep{Fang2016}. Numerical simulations have extensively shown that the presence of a mixing or conduction layer regulates the reduction of the temperature of the adiabatically-shocked hot bubble when in contact with the nebular, optically-emitting shell, ultimately producing the observed soft X-ray emission. However, it is still not clear whether thermal conductivity or hydrodynamical mixing dominate this procedure because they produce similar X-ray luminosities \citep{Toala2016}. In order to understand further the physics behind the production of soft X-ray-emitting gas from wind-blown bubbles we require synthetic UV and X-ray spectra to study different line ratios formed under different physical conditions (thermal conduction or hydrodynamical mixing).

\subsection{Mixing layer in other WRPNe}

We extended our {\sc cloudy} calculations to other [WR]-type stars in order to explore the usefulness of the C\,{\sc iv} and N\,{\sc v} emission lines predicting the presence of a mixing layer and, as a consequence, the presence of an X-ray-emitting hot bubble. For this, we used stellar atmosphere models produced with {\sc powr} for the CSPNe of NGC\,5189, NGC\,6905 and NGC\,6369 which are representative of [WO1], [WO2] and [WO3]-type stars \citep[][and Toal\'{a} et al. in prep.]{GG2022}, respectively. In addition, templates of [WC4] and [WC5] were also retrieved from the {\sc powr} database. Similar calculations as those presented in Section~\ref{sec:mix_ngc1501} were computed but for the different stellar atmosphere models by keeping the nebular abundances to those of NGC\,1501. The results are illustrated in Fig.~\ref{fig:lineas_cloudy2}.

Within model limitations, the intensity variation of the C\,{\sc iv}, N\,{\sc v} and O\,{\sc vi} lines presented in Fig.~\ref{fig:lineas_cloudy} and \ref{fig:lineas_cloudy2} can be used to predict the presence of a mixing layer in WRPN harbouring different [WR]-types as late as [WC5]. Our models confirm that the O\,{\sc vi} lines can be practically used in all cases to study the presence of a mixing layer. However, care must be payed for compact WRPNe harbouring [WO1] stars (see Fig.~\ref{fig:lineas_cloudy2}). The hottest cases, [WO1] and [WO2], exhibit a maxima in the evolution of the C\,{\sc iv} lines. These are produced by intermediate size nebulae and only reflects the dominance of C$^{4+}$ (not shown here), which has a ionization potential of $\sim$65~eV.

Similarly to what is found for the [WO4] model (Fig.~\ref{fig:lineas_cloudy}), the C\,{\sc iv} and N\,{\sc v} lines will be always brighter in UV wavelengths given the atomic properties of these resonant lines \citep{Wiese1996}. However, the present works demonstrated that high-resolution, high-S/N spectroscopic observations can be also used to disclose the presence of mixing regions using optical data in WRPNe, but such observations are not numerous in the literature. There are some works addressing the presence of these lines in UV wavelengths. Examples are listed in Table~\ref{tab:uv_fluxes} for WRPNe harbouring [WO1]--[WO4]-type stars. 

This table shows that in addition to NGC\,1501, the are other two WRPNe harbouring [WO4]-type CPSN that also exhibit UV fluxes that surpass those predicted by the photoionization models and subsequently the presence of a mixing region is strongly suggested. NGC\,5315 and NGC\,6751 have been reported to have C\,{\sc iv} UV lines almost two orders of magnitude larger that our predictions for a [WO4] model (see Fig.~\ref{fig:lineas_cloudy}). This is also the case for the N\,{\sc v} lines of NGC\,5315. Serendipitous {\it Chandra} observations of NGC\,5315 have placed this WRPNe as the brightest X-ray-emitting PNe \citep[e.g.,][and Montez, Toal\'{a} et al. in prep.]{Kastner2008}. On the other hand, NGC\,6751 clearly indicate the presence of a bright mixing layer and thus the presence of a bright X-ray-emitting bubble. Unfortunately, it has not been observed in X-rays.

There are two WRPNe listed in Table~\ref{tab:uv_fluxes} that are not conclusive. The UV flux of the N\,{\sc v} lines from the compact WRPN Hen\,2-55, which harbours a [WO3]-type CSPN, suggest the presence of a mixing layer. But the estimated flux from the C\,{\sc iv} lines seem to be consistent with photoionization. On the other hand, comparison of the C\,{\sc iv} emission lines from NGC\,2867 with results from [WO2]-type models are also consistent with photoionization, but we note that there is not report of the N\,{\sc v} in the literature for this WRPN.

The other three sources listed in Table~\ref{tab:uv_fluxes}, namely NGC\,2452, NGC\,5189 and NGC\,6905, have UV fluxes that are consistent with the presence of mixing layers. In particular, NGC\,2452 has C\,{\sc iv} fluxes that are consistent with photoionization given the early type of its CSPN, [WO1], but the N\,{\sc v} emission is above {\sc cloudy} predictions. Predictions for the extended WRPNe NGC\,5189 and NGC\,6905 (not shown here) are also consistent with the presence of mixing layers. In particular, NGC\,5189 is the most extended PNe with a hot bubble that seems to have been powered by a born-again event, which produced the C overabundance of the X-ray-emitting material \citep{Toala2019a}.

Our models confirm that the mixing layer can be also unveiled by the presence of C\,{\sc iv} and N\,{\sc v} lines even for high-temperature [WR]-type CSPN. We corroborated that WRPNe detected with X-ray observatories also display the presence of a mixing region revealed by these UV lines. Finally, we predict that NGC\,2452, NGC\,6751 and NGC\,6905 will be detected by future X-ray observations.

\section{Conclusions}
We presented a thorough photoionization model of the WRPN NGC\,1501 able to reproduce the observed optical, IR and radio observations, and its physical properties. To achieve this model, we study the stellar atmosphere of its CSPN by means of the state-of-the-art {\sc powr} code. Our {\sc cloudy} photoionization model of NGC\,1501 allowed us to estimated a total mass of the nebula of $\sim$0.22~M$_\odot$, which includes $\sim$0.21~M$_\odot$ ionized gas and 8.9$\times$10$^{-4}$~M$_\odot$ of C-rich dust. Taking into account a current mass of 0.60$^{+0.06}_{-0.02}$~M$_\odot$ for the CSPN, we estimated an initial mass of 0.80--0.88~M$_\odot$, which is consistent with the abundances determination when compared with stellar evolution models predictions.

Theoretical models predict that the soft X-ray emission from wind-blown bubbles is produced and regulated by a conductive or mixing layer. It has been suggested that it can be unveil by C\,{\sc iv}, N\,{\sc v} and O\,{\sc vi} emission lines. We demonstrated that the C\,{\sc iv} and N\,{\sc v} lines detected in the optical and UV from NGC\,1501 are produced in a mixing layer between the hot X-ray-emitting bubble and the ionized gas. The mixing layer is in pressure equillibrium with the hot bubble, confirming that they are produced by the same physical mechanisms. Our calculations render NGC\,1501 only the second PN for which this phenomenon has been characterized, after NGC\,6543.

We extended our photoionization models to show that even H-deficient hot [WO] stars cannot produce C\,{\sc iv}, N\,{\sc v} and O\,{\sc vi} by photoionization and, thus, they can be use to unveil the presence of conductive or mixing layers by optical and UV observations. WRPNe that have been previously found to be X-ray-bright exhibit UV fluxes that confirm the presence of mixing or conductive layer. We suggest that NGC\,2452, NGC\,6751 and NGC\,6905 are very likely to be detected by X-ray observatories.

Future radiation-hydrodynamic numerical simulations including non-equilibrium ionization processes in combination with high S/N detections of the C\,{\sc iv}, N\,{\sc v} and O\,{\sc vi} will help disclosing the physics behind the production of soft X-ray emission from PNe, which will paved the way for future X-ray missions.

\section*{Acknowledgements}

The authors are thankful to the anonymous referee for comments and suggestions that improved the analysis and presentation of the results. G.R. and E.S. acknowledge support from Consejo Nacional de Ciencia y Tecnolog\'{i}a (CONACyT) for student scholarship. J.A.T. is funded by UNAM DGAPA PAPIIT project IA101622, the Marcos Moshinsky Fundation (Mexico) and the Visiting-Incoming
programme of the IAA-CSIC through the Centro de Excelencia Severo Ochoa. M.A.G. acknowledges support of grant PGC 2018-102184-B-I00 of the Ministerio de Educaci\'{o}n, Innovaci\'{o}n y Universidades cofunded with FEDER funds. G.R.-L. acknowledges support from CONACyT (grant 263373) and PRODEP (Mexico). L.S. acknowledges support from PAPIIT UNAM grant IN110122. This work makes use of {\sc iraf}, distributed by the National Optical Astronomy Observatory, which is operated by the Association of Universities for Research in Astronomy under cooperative agreement with the National Science Foundation. This work makes use {\it Spitzer} and {\it WISE} IR observations. The {\it Spitzer} Space Telescope was operated by the Jet Propulsion Laboratory, California Institute of Technology under a contract with NASA. Support for this work was provided by NASA through an award issued by JPL/Caltech. {\it WISE} is a joint project of the University of California (Los Angeles, USA) and the JPL/Caltech, funded by NASA. This work has made extensive use of NASA's Astrophysics Data System (ADS).

\section*{Data availability}
The data underlying this work are available in the article.
The reduced observations files will be shared on reasonable request to the first author.


\end{document}